\documentclass[10pt,aps,prd,notitlepage,preprintnumbers]{revtex4-1}
\usepackage{graphicx}
\usepackage{slashed}
\usepackage{xcolor}
\usepackage{amsmath}
\usepackage{amssymb}
\usepackage{amsfonts}
\usepackage[bookmarks=false,colorlinks, linkcolor=green]{hyperref}
\newcommand{\ep}{\epsilon}
\newcommand{\la}{\lambda}
\newcommand{\de}{\delta}
\newcommand{\ga}{\gamma}
\newcommand{\al}{\alpha}
\newcommand{\be}{\beta}
\newcommand{\sig}{\sigma}
\newcommand{\s}[1]{\slashed{#1}}

\newcommand{\no}{\notag\\}
\newcommand{\lrb}[1]{\left({#1}\right)}
\newcommand{\msb}{\overline{\text{MS}}}
\newcommand{\gev}{\text{GeV}}

\begin{abstract}
The azimuthal asymmetry of heavy quarks production on double polarized
proton-proton and proton-antiproton colliders are studied in this work
at next-to-leading order in $\alpha_s$, with some details included. The
purpose is to see whether
the effect of extracted transversity
distribution functions can be seen on present and near future colliders.
All analytic one-loop hard coefficients are given. Numerical results
for the asymmetry on proton-(anti)proton colliders are presented.
\end{abstract}

\begin{document}
\title{Transversity and heavy quark production on hadron colliders}
\author{G.P. Zhang}
\email[]{zgp-phys@pku.edu.cn}
\affiliation{Department of physics, Yunnan University, Kunming, Yunnan 650091, China}
\maketitle

\section{Introduction}
Transversity distribution function of quark is one of three twist-2 parton distribution functions(PDFs), which reflects
the spin structure of proton\cite{Ralston:1979ys,Jaffe:1991kp,Jaffe:1991ra,Cortes:1991ja}.
Compared with other two PDFs,
the extraction of transversity PDF is much more difficult. Due to its chiral-odd
nature, it must convolute with another chiral-odd distribution or fragmentation function to
form an observable.
Through many years of efforts, now the transversity PDFs in valence region are available. There are two independent extraction formalisms in
literature: One is based on transverse momentum dependent(TMD) factorization
formalism, for which one has to determine Collins function at the same
time\cite{Anselmino:2013vqa,Kang:2015msa,Lin:2017stx}; Another
one is based on collinear formalism, with Di-hadron fragmentation functions
as input\cite{Radici:2018iag}. Within uncertainty range the results of these two formalisms are in agreement. In both schemes sea transversity cannot be determined at present.
On the other hand, double spin asymmetry(DSA), including double polarized Drell-Yan,
single jet or photon production
(see e.g.,\cite{Jaffe:1991ra,Barone:2005cr,Shimizu:2005fp,Kawamura:2007ze,deFlorian:2017ogw,Soffer:2002tf,Mukherjee:2003pf,Mukherjee:2005rw})
has been proposed for a long time to extract transversity distributions.
Since sea transversity is expected to be small, the resulting DSAs on proton-proton
colliders, such as RHIC\cite{RHIC-spin:2013woa}, are usually very small. However,
as long as the production rate is high, it is not hopeless to see the effect of
transversity PDFs. Besides lepton pair, jet and photon, the heavy
quark(such as bottom) production rate on RHIC is also very high, which is of order $10^3 pb$,
thus may provide
some opportunities to see the effect of sea transversity or give a bound to sea
transversity. If polarized anti-proton beam is available in future, as proposed by
PAX collaboration at GSI\cite{Barone:2005pu,GSI:2004dza}, valence transversity PDFs will be
detected directly through double polarized Drell-Yan process.
As an important background to polarized Drell-Yan, heavy quark production
has to be known. In this work, we study the production rate of single inclusive heavy quark
in hadron-hadron collision
with the initial two hadrons transversely polarized. The result may
help to check extracted transversity PDFs.

The structure of this paper is as follows: In Sect.II, we make clear our formalism;
in Sect.III,
we present virtual and real one-loop corrections and the subtracted result. Some
details for the reduction and calculation scheme of real correction will be
given;in Sect.IV,
the numerical results on proton-(anti)proton colliders are described and
Sect.V is our summary.

\section{Formalism and tree level result}
\begin{figure}
\includegraphics[scale=0.5]{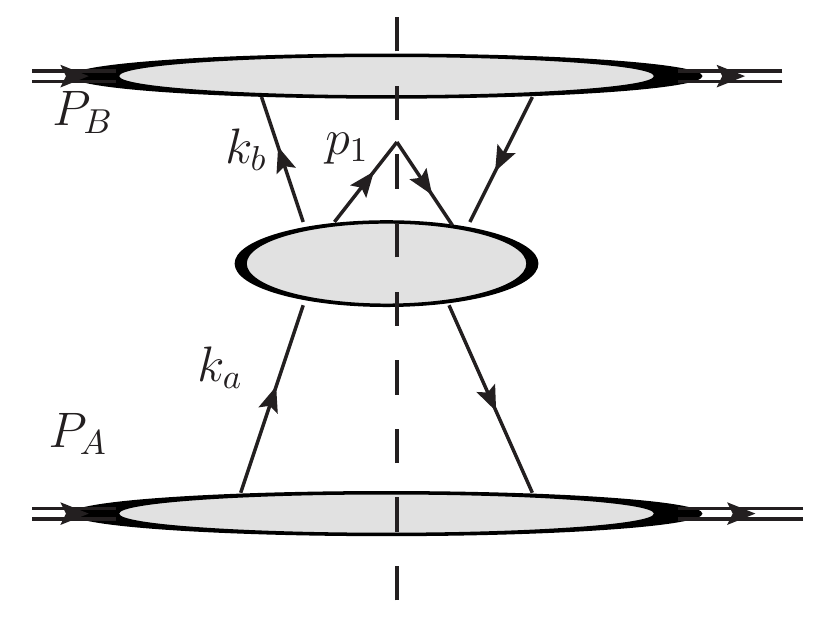}
\caption{Leading region for heavy quark production. The central bubble represents
hard region, and upper and lower bubbles represent collinear regions.}
\label{fig:leading_region}
\end{figure}
The process we want to study is
\begin{align}
h_A(P_A,s_{a\perp})+h_B(P_B,s_{b\perp})\rightarrow Q(p_1)+X,
\end{align}
where $h_A$, $h_B$ with momentum $P_A, P_B$ are two transversely polarized hadrons, which can be proton or anti-proton in our case. $s_{a\perp}$, $s_{b\perp}$ are corresponding spin vectors, which are perpendicular to momenta $\vec{P}_A$,  $\vec{P}_B$ in the center of mass system(cms) of initial hadrons. $Q(p_1)$ is the detected heavy quark(bottom or charm), with momentum $p_1$. Our calculation will
be performed in cms of initial hadrons.

For heavy quark(charm or bottom) production, the quark mass is much greater than
the typical hadron scale of QCD, i.e. $\Lambda_{QCD}$,
thus perturbative calculation is allowed. Besides, we require the detected heavy
quark has a large transverse momentum $p_{1\perp}$, which is usually larger than
quark mass. By combining these two hard scales together,
the typical hard scale for this process could be taken as the transverse energy defined as $E_{1\perp}=\sqrt{\vec{p}_{1\perp}^2 +m^2}$. In the following, we
will give the leading power cross sections under the expansion in
$\Lambda_{QCD}/E_{1\perp}$, which is called twist-2 contribution.

The calculation of twist-2 cross section for heavy quark production is very
standard. Unpolarized cross section at next-to-leading order(NLO) in strong
coupling $\al_s$ expansion has been calculated long before\cite{Nason:1989zy,Beenakker:1988bq,Beenakker:1990maa}.
So far, even next-to-next-to-leading order(NNLO) result is available(see \cite{Catani:2019hip} for example).
Here we just give a simple derivation of the factorization formula. More formal discussions
can be found in \cite{Collins:2011zzd} and reference therein.

Throughout the paper, we work in cms of initial hadrons and
use light-cone coordinates. For any four vector $a^\mu$,
its components are denoted by $a^\mu=(a^+,a^-,a_\perp^\mu)$, with
$a^\pm=(a^0\pm a^3)/\sqrt{2}$. In addition, $\vec{P}_A$ is along $+z$ axis.
Under high energy limit $E_{1\perp}\gg \Lambda_{QCD}$, the hadron masses can be ignored and then $P_A^\mu\simeq(P_A^+,0,0)$, $P_B^\mu\simeq(0,P_B^-,0)$. For
convenience we introduce a transverse metric:
\begin{align}
g_\perp^{\mu\nu}=g^{\mu\nu}-\frac{P_A^\mu P_B^\nu + P_A^\nu P_B^\mu}{P_A\cdot P_B},
\end{align}
then the transverse components of any vector $a^\mu$ is given by
$a_\perp^\mu=g_\perp^{\mu\nu}a_\nu$.

Under high energy limit,$E_{1\perp}\gg \Lambda_{QCD}$,
collinear partons give leading power or twist-2 contribution. The leading region
for quark contribution is shown in Fig.\ref{fig:leading_region}, where the momenta of partons $k_a,k_b$
are collinear to external momenta $P_A, P_B$, respectively. That is,
\begin{align}
k_a^\mu=(k_a^+,k_a^-,k_{a\perp}^\mu)\simeq E_{1\perp}(1,\la^2,\la),
k_b^\mu\simeq E_{1\perp}(\la^2,1,\la),\la\simeq \Lambda_{QCD}/E_{1\perp}\ll 1.
\end{align}
According to Fig.\ref{fig:leading_region}, the cross section is written as
\begin{align}
d\sig(s_{a\perp},s_{b\perp})=&\frac{1}{2S}\int\frac{d^{n-1}p_1}{(2\pi)^{n-1} 2E_1}\int d^nk_a d^n k_b
\int\frac{d^n\xi_a}{(2\pi)^n}\frac{d^n\xi_b}{(2\pi)^n}e^{ik_a\cdot \xi_a}
e^{ik_b\cdot \xi_b}H_{ij}^{mn}(k_a,k_b,p_1)\no
&\langle P_As_a|\bar{\psi}_j(0)\psi_i(\xi_a)|P_A s_a\rangle \langle P_Bs_b|\psi_n(0)\bar{\psi}_m(\xi_b)|P_B s_b\rangle, S=(P_A+P_B)^2,
\end{align}
where $ij,mn$ are color and Dirac indices of partons, and $H_{ij}^{mn}$ is the hard part in which inner propagators are far off-shell.
Since $k_{a\perp},k_{b\perp}$ are much smaller than $E_{1\perp}$ in $H_{ij}^{mn}$,
they can be ignored at leading power level. This gives twist-2 hard coefficients.
After this approximation, $k_{a\perp},k_a^-$ and $k_{b\perp},k_b^+$ can be integrated over in the correlation functions. Then,
\begin{align}
d\sig(s_{a\perp},s_{b\perp})
=&\frac{1}{2S}\int\frac{d^{n-1}p_1}{(2\pi)^{n-1} 2E_1}\int dk_a^+  dk_b^- H_{ij}^{mn}(k_a^+,k_b^-,p_1)\no
&\int \frac{d\xi_a^-}{2\pi}e^{ik_a^+ \xi_a^-}\langle P_As_a|\bar{\psi}_j(0)\psi_i(\xi_a^-)|P_A s_a\rangle
\int \frac{d\xi_b^+}{2\pi}
e^{ik_b^-\xi_b^+}\langle P_Bs_b|\psi_n(0)\bar{\psi}_m(\xi_b^+)|P_B s_b\rangle.
\end{align}
The correlation functions defined on the light-cone can be projected to PDFs\cite{Jaffe:1991kp} as
\begin{align}
\int\frac{d\xi^-}{2\pi}e^{-i\xi^- x P_A^+}\langle P_A s|\bar{\psi}_j(\xi^-)\psi_i(0)|P_A s\rangle
=\frac{1}{2N_c}\de_{ij}\Big[\ga_5\s{s}_\perp\ga^- h_1(x)
+\ga^- f_1(x)\Big]_{ij},
\label{eq:quark_PDF}
\end{align}
which is for quark PDF and
\begin{align}
\int\frac{d\xi^-}{2\pi}e^{-i\xi^- xP_A^+}\langle P_A s|\psi_i(\xi^-)\bar{\psi}_j(0)|P_A s\rangle
=\frac{1}{2N_c}\de_{ij}\Big[\ga_5\s{s}_\perp\ga^- \bar{h}_1(x)
+\ga^- \bar{f}_1(x)\Big]_{ij},
\label{eq:antiquark_PDF}
\end{align}
which is for anti-quark PDF. Since we only consider transverse spin effect in this
work, we have ignored the longitudinal components of spin vectors in the two equations above.

Apparently, chiral-odd PDF $h_1$ will not combine with chiral-even PDF $f_1$ to
give a nonzero cross section since light quark is taken as massless. Our final
factorization formula is
\begin{align}
\frac{d\sig(s_{a\perp},s_{b\perp})}{dy d^2 p_{1\perp}}=&
\frac{d\sig_{unp}}{dy d^2 p_{1\perp}}+\frac{d\Delta\sig}{dy d^2 p_{1\perp}},
\end{align}
and
\begin{align}
\frac{d\sig_{unp}}{dy d^2 p_{1\perp}}=\int dx_a dx_b
f_1(x_a,\mu)\bar{f}_1(x_b,\mu)\frac{d\hat{\sig}_{unp}(k_a^+,k_b^-,p_1)}{dy d^2 p_{1\perp}},\
\frac{d\Delta\sig}{dy d^2 p_{1\perp}}=\int dx_a dx_b
h_1(x_a,\mu)\bar{h}_1(x_b,\mu)\frac{d\Delta\hat{\sig}(k_a^+,k_b^-,p_1)}{dy d^2 p_{1\perp}},
\label{eq:hadron_cross}
\end{align}
where $d\hat{\sig}_{unp}$ is spin independent partonic cross section and
$d\Delta\hat{\sig}$ is spin dependent part which is proportional to $s_{a\perp}$
and $s_{b\perp}$. The explicit formulas for these two partonic cross sections
are
\begin{align}
\frac{d\hat{\sig}_{unp}}{dy d^2 p_{1\perp}}=& \frac{1}{8(2\pi)^3}
\lrb{\frac{1}{2N_c}}^2 H_{ij}^{mn}(k_a^+,k_b^-,p_1)
[\de_{ij}\ga^-_{ij}][\de_{mn}\ga^+_{mn}],\no
\frac{d\Delta\hat{\sig}}{dy d^2 p_{1\perp}}=& \frac{1}{8(2\pi)^3}
\lrb{\frac{1}{2N_c}}^2 H_{ij}^{mn}(k_a^+,k_b^-,p_1)
\de_{ij}[\ga_5\s{s}_{a\perp}\ga^-]_{ij}
\de_{mn}[\ga_5\s{s}_{b\perp}\ga^+]_{mn}.
\label{eq:fac_formula}
\end{align}
Note that $H_{ij}^{mn}$ is obtained by removing parton propagators connecting
hard part and collinear(or jet) part in Fig.\ref{fig:leading_region}. $\de_{ij},\de_{mn}$ are color matrices in color space.

Next we analyze the spin structure of polarized partonic cross
section $d\Delta\hat{\sig}$. Since the two spin vectors are transverse and there
is only one transverse momentum $p_{1\perp}$ in this process,  we must have
two structure functions $F_{1,2}$, which are defined as
\begin{align}
\frac{d\Delta\hat{\sig}}{dy d^2 p_{1\perp}}\equiv & s_{a\perp}^\al s_{b\perp}^\be
W_\perp^{\al\be}(k_a^+,k_b^-,p_1)
=s_{a\perp}^\al s_{b\perp}^\be
\Big[
F_1\left(g_\perp^{\al\be}-\frac{(n-2)p_{1\perp}^\al p_{1\perp}^\be}{p_{1\perp}^2}\right)+F_2 g_\perp^{\al\be}
\Big],\ \ p_{1\perp}^2=-\vec{p}_{1\perp}^2<0.
\end{align}
There is no anti-symmetric tensor $\ep^{\mu\nu\rho\tau}$ in the tensor decomposition, because in the Dirac trace $\ga_5$ appears in pair. Now $F_{1,2}$
do not contain spin vectors any more, they can be calculated in the same way as
that for unpolarized partonic cross section. Since both ultra-violate(UV) and infra-red(IR) divergences will appear in $F_{1,2}$ beyond tree level,
we make use of dimensional regularization for these two kinds of
divergences. The space-time dimension is $n=4-\ep$. Unless declared explicitly, $1/\ep$ represents UV or IR poles. After renormalization and subtraction of IR
divergences we take the limit $\ep\rightarrow 0$.

Now the spin dependence is clear, and the azimuthal angle dependence can be obtained. We consider a special case, in which
$\vec{s}_{a\perp}\parallel \vec{s}_{b\perp}$.
Other spin configurations can be considered, but one cannot get more information. Suppose both
$\vec{s}_{a\perp}$ and $\vec{s}_{b\perp}$ are along
$+x$-axis and the azimuthal angle of $\vec{p}_{1\perp}$ relative to $+x$-axis
is $\phi$. We have
\begin{align}
\frac{d\Delta\hat{\sig}}{dy d^2 p_{1\perp}}=&|\vec{s}_{a\perp}||\vec{s}_{b\perp}|
\Big[
(1-\frac{\ep}{2})F_1(p_{1\perp}^2)\cos{2\phi}-\frac{\ep}{2}F_1(p_{1\perp}^2)
-F_2(p_{1\perp}^2)\Big].
\end{align}
What we are interested in is the $\cos{2\phi}$ distribution. Especially,
in our calculation we find that at one-loop level $\ep F_1$
and $F_2$ can be ignored, because they are $O(\ep)$ after renormalization and
collinear subtraction. In the following we will only
consider the contribution of $F_1$. By substituting partonic cross section into
eq.(\ref{eq:hadron_cross}), the hadron cross section is
\begin{align}
\frac{d\sig(s_{a\perp},s_{b\perp})}{dy d^2 p_{1\perp}}=&
\int dx_a dx_b\Big[
f_1(x_a,\mu)\bar{f}_1(x_b,\mu)\frac{d\hat{\sig}_{unp}(k_a^+,k_b^-,p_1)}{dy d^2 p_{1\perp}}
+\cos{2\phi}|\vec{s}_{a\perp}||\vec{s}_{b\perp}|
h_1(x_a,\mu)\bar{h}_1(x_b,\mu)F_1(k_a^+,k_b^-,p_1)
\Big].
\end{align}
Now the double spin asymmetry(DSA) is defined as
\begin{align}
A_{TT}=\frac{\int_0^{2\pi} d\phi \cos{2\phi} \Big[d\sig(s_{a\perp},s_{b\perp})-d\sig(s_{a\perp},-s_{b\perp})
\Big]}{
\int_0^{2\pi} d\phi
\Big[d\sig(s_{a\perp},s_{b\perp})+d\sig(s_{a\perp},-s_{b\perp})
\Big]},
\label{eq:ATT_def}
\end{align}
where the azimuthal angle of detected heavy quark, $\phi$, is integrated over.
Expressed in partonic quantities, $A_{TT}$ is
\begin{align}
A_{TT}=& |\vec{s}_{a\perp}||\vec{s}_{b\perp}|\frac{\pi}{2}
\frac{\int dx_a dx_b h_1(x_a,\mu)\bar{h}_1(x_b,\mu)F_1}{
\int dx_a dx_b f_1(x_a,\mu)\bar{f}_1(x_b,\mu)\frac{d\hat{\sig}_{unp}}{
dy d\vec{p}_1^2}}.
\end{align}
Note that we have used the relation
$\pi d\sig_{unp}/dy d^2 p_{1\perp}=d\sig_{unp}/dy d\vec{p}^2_{1\perp}$ for unpolarized cross
section.

For the calculation of partonic cross sections, we use the Lorentz invariants
defined in \cite{Nason:1989zy}, that is,
\begin{align}
\tau_1=\frac{k_a\cdot p_1}{k_a\cdot k_b},\ \tau_2=\frac{k_b\cdot p_1}{k_a\cdot k_b},
\ \rho=\frac{4m^2}{s},\ \tau_x\equiv 1-\tau_1-\tau_2,\ s=(k_a+k_b)^2.
\end{align}

Generally, partonic structure function
$F_1$ can be organized in a neat way as done in \cite{Nason:1989zy}, i.e.,
\begin{align}
F_1=& \Delta H_d\de(\tau_x)+ \Delta H_p \lrb{\frac{1}{\tau_x}}_+
+ \Delta H_l \lrb{\frac{\ln\tau_x}{\tau_x}}_+,\no
\Delta H_d
=&\frac{\al_s^2}{s^2}\Big[\Delta h_d^{(0)}
+\frac{\al_s}{2\pi}\Delta h_d^{(1)}+\cdots\Big],\
\Delta H_p=\frac{\al_s^2}{s^2}\Big[\frac{\al_s}{2\pi}\Delta h_p^{(1)}+\cdots\Big],\
\Delta H_l=\frac{\al_s^2}{s^2}\Big[\frac{\al_s}{2\pi}
\Delta h_l^{(1)}+\cdots\Big].
\end{align}
The plus function is the standard one\cite{Nason:1989zy}. Tree level result is
given by the process $q(k_a)\bar{q}(k_b)\rightarrow Q(p_1)+\bar{Q}$, and we get
\begin{align}
\Delta h_d^{(0)}=&-\frac{C_F}{2N_c}(\rho-4\tau_1(1-\tau_1))\Big[
1+\frac{\ep}{2}+\frac{\ep^2}{4}+O(\ep^3)\Big].
\end{align}
For $F_2$ we get
\begin{align}
F_2=\frac{\al_s^2}{s^2}\frac{C_F}{2N_c}\frac{\ep}{2}[\rho+2-4\tau_1(1-\tau_1)]
\de(\tau_x).
\end{align}
As we see, $F_2$ is $O(\ep)$. To one-loop level, $F_2$ can have a finite
part, but it has been checked explicitly that the finite part is removed by
renormalization and collinear subtraction. The resulting $F_2$ is still $O(\ep)$
and can be ignored. In the following we will not discuss $F_2$ any more.

For unpolarized partonic cross section, similar hard coefficients are defined
as follows
\begin{align}
\frac{d\hat{\sig}^{unp}}{dy d^2 p_{1\perp}}=&
H_d\de(\tau_x)+ H_p \lrb{\frac{1}{\tau_x}}_+
+H_l \lrb{\frac{\ln\tau_x}{\tau_x}}_+.
\end{align}
Compared with the decomposition of $F_1$, all $\Delta H_{d,p,l}$ are replaced
by $H_{d,p,l}$. The expansion of $H_{d,p,l}$ in terms of $h_{d,p,l}$ is
the same as that for $\Delta H_{d,p,l}$.
Tree level unpolarized result from $q\bar{q}$ scattering is
\begin{align}
h_d^{(0)}\Big|_{q\bar{q}}=&\frac{C_F}{2N_c}
\Big[2+\rho-4\tau_1(1-\tau_1)-\ep +O(\ep^3)\Big].
\end{align}

\section{One-loop correction}

\subsection{One-loop virtual correction}
\begin{figure}
\includegraphics[scale=0.5]{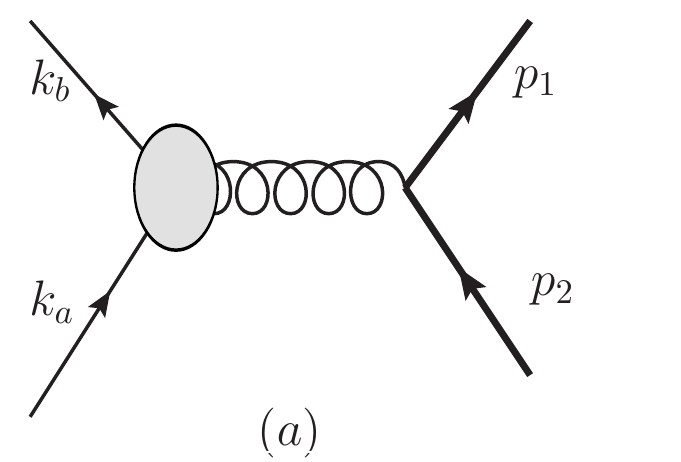}
\includegraphics[scale=0.5]{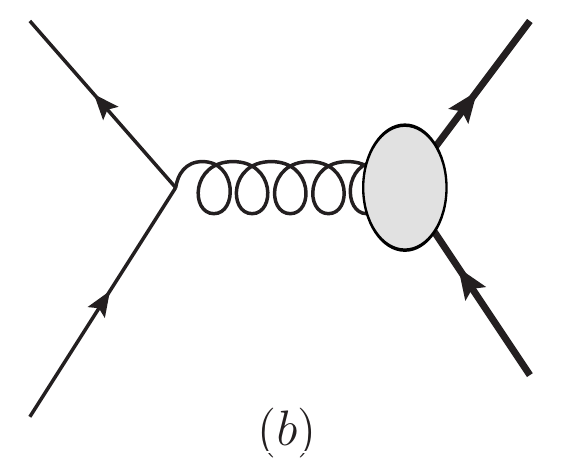}
\includegraphics[scale=0.5]{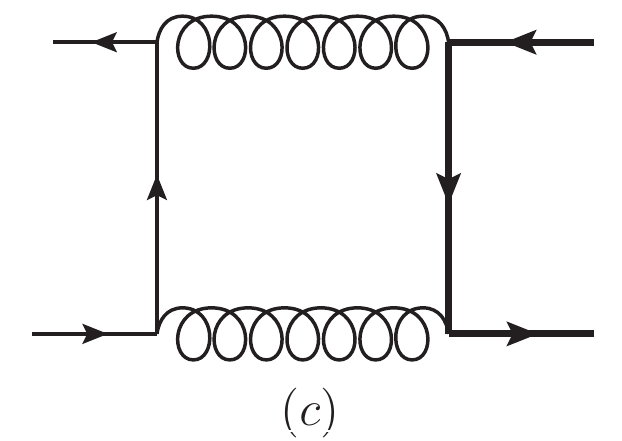}
\includegraphics[scale=0.5]{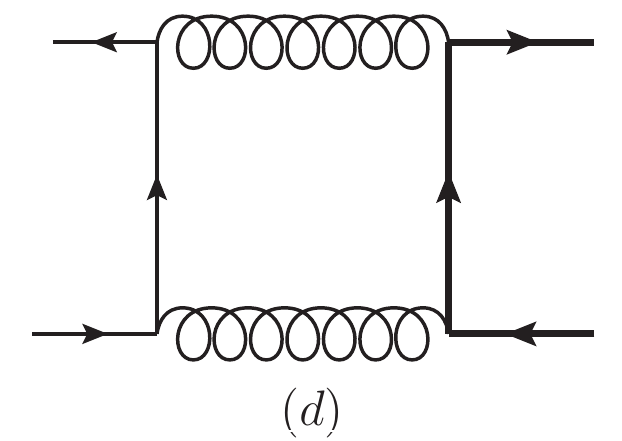}
\includegraphics[scale=0.5]{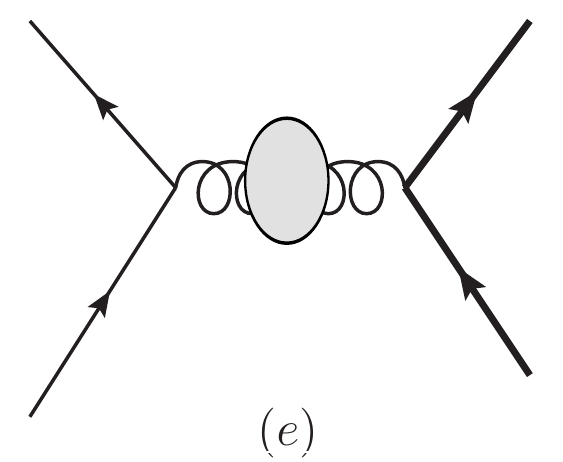}
\caption{Virtual corrections to the amplitude, where the bubbles in (a,b)
represent vertex insertion and the bubble in (e) represents the insertion of
gluon self-energy in Feynman gauge.
Self-energy insertion to external legs is not shown but
included in the calculation. Thick lines represent heavy quarks. }
\label{fig:virtual}
\end{figure}

All diagrams appearing in virtual correction are shown in Fig.\ref{fig:virtual}. Self-energy insertions to
external lines are trivial and not shown, but included in our calculation.
Calculating these diagrams is very straightforward. The tensor integrals
are reduced by FIRE\cite{Smirnov:2008iw}. Resulting scalar integrals are standard and can be found for example in \cite{Beenakker:1988bq,Ellis:2007qk}.
We note that by integration by part relations(IBPs) for integral reduction,
the divergent pole $1/\ep$ may transfer from tensor or scalar integrals to
reduced coefficients. In order to get finite result, the resulting scalar
integrals should be expanded to higher orders in $\ep$. In our case, we find
it necessary to expand bubble and tadpole integrals to $O(\ep)$. The bubble
and tadpole integrals are listed in Appendix.\ref{sec:virtual_integrals}.

There are UV, soft and collinear divergences in virtual correction. But a simple
analysis indicates these divergences are independent of the polarization status
of initial partons. This is confirmed by our explicit calculation.
The divergent hard coefficients from Fig.\ref{fig:virtual} are
\begin{align}
\{h_d^{(1)},\Delta h_d^{(1)}\}=& \{h_d^{(0)},\Delta h_d^{(0)}\}
16\pi R_\ep\lrb{\frac{\mu^2}{m^2}}^{\ep/2}
\Big[C_F \Delta G_1+C_A \Delta G_2+ \Delta G_3\Big],\ \
R_\ep\equiv\frac{(4\pi)^{\ep/2}}{8\pi\Gamma(1-\ep/2)}, \no
\Delta G_1=&-\frac{4}{\ep^2}-\frac{2}{\ep}\Big[
\frac{2-\rho}{2\sqrt{1-\rho}}\ln\frac{1-\sqrt{1-\rho}}{1+\sqrt{1-\rho}}
+1+\ln\frac{\rho}{4}+4\ln\frac{1-\tau_1}{\tau_1}\Big],\no
\Delta G_2=&\frac{1}{\ep}
\Big[
\frac{2-\rho}{2\sqrt{1-\rho}}\ln\frac{1-\sqrt{1-\rho}}{1+\sqrt{1+\rho}}
+\frac{11}{3}-\ln\frac{\rho}{4}+2\ln\frac{(1-\tau_1)^2}{\tau_1}\Big]
,\no
\Delta G_3=&-\frac{2(2+n_F)}{3\ep},
\label{eq:virtual_correction}
\end{align}
where $n_{F}$ is the number of light fermion flavors. Throughout this paper,
it is equal to 3. Both charm and bottom are included in the fermion loops appearing
in gluon self-energy, i.e., Fig.\ref{fig:virtual}(e).
Wave function renormalization and UV counter terms(c.t.) give
\begin{align}
\{h_d^{(1)},\Delta h_d^{(1)}\}\Big|_{\text{wave+c.t.}}=&
\{h_d^{(0)},\Delta h_d^{(0)}\}
\Big[\frac{2\pi}{\al_s}\lrb{2(Z_2-1)+2(Z_2^{(m)}-1)}
-16\pi R_\ep\frac{11 C_A +6 C_F -2(2+n_F)}{3\ep}\Big],
\end{align}
The factor $2\pi/\al_s$ is caused by the definition of $h_d^{(1)}$ or $\Delta h_d^{(1)}$.
Wave function renormalization constants for massless and massive fermions are
\begin{align}
Z_2-1=\frac{g_s^2 C_F}{16\pi^2}
\lrb{\frac{2}{\ep_{IR}}-\ga_E+\ln 4\pi}
,\
Z_2^m-1
=-\frac{g_s^2 C_F}{16\pi^2}\Big[
2\lrb{\frac{2}{\ep_{IR}}-\ga_E+\ln 4\pi}
+3\ln\frac{\mu^2}{m^2}+4\Big].
\label{eq:wave_function}
\end{align}
For convenience, we have removed UV poles by using $\msb$ renormalization scheme.
In addition, mass renormalization for massive quark is done in pole mass scheme.
The renormalized mass $m$ is a physical mass and does not depend on renormalization
scale. For simplicity, we do not plan to show the finite corrections in this paper,
but leave them in the mathematica files, which can be obtained from author if required.

\subsection{One-loop Real correction}
\begin{figure}
\begin{flushleft}
\includegraphics[width=0.12\textwidth]{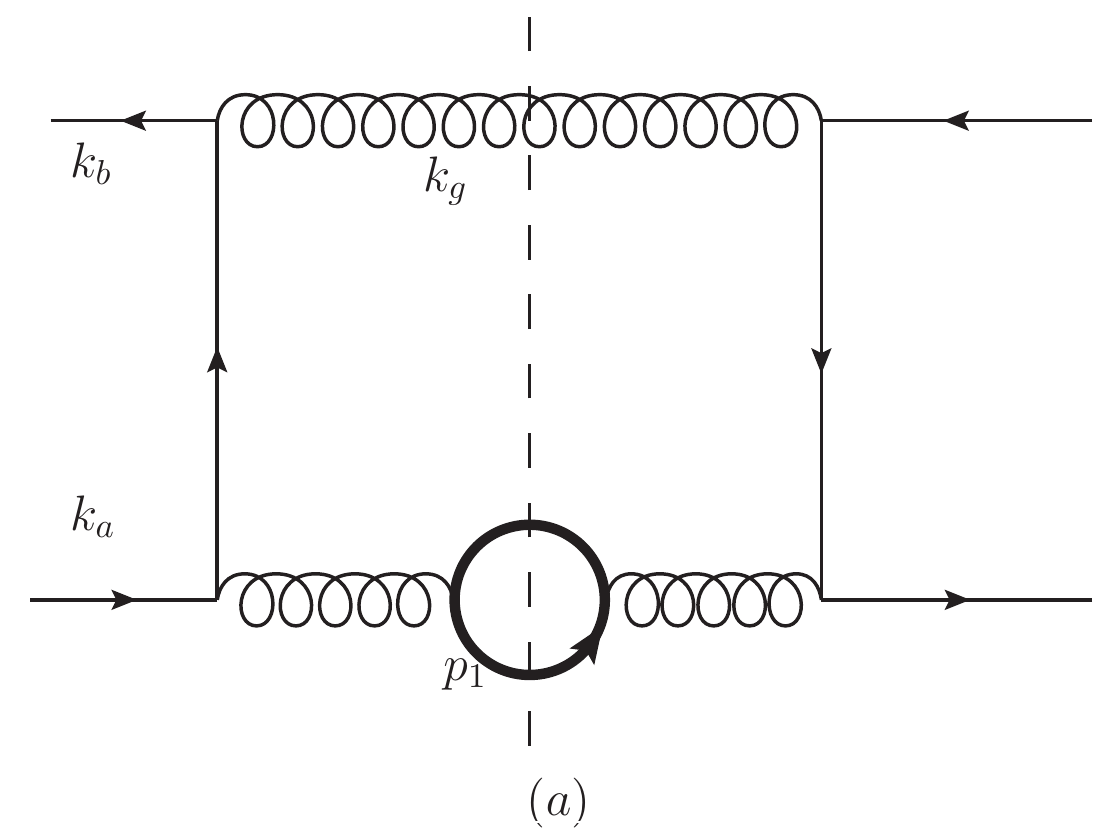}
\includegraphics[width=0.12\textwidth]{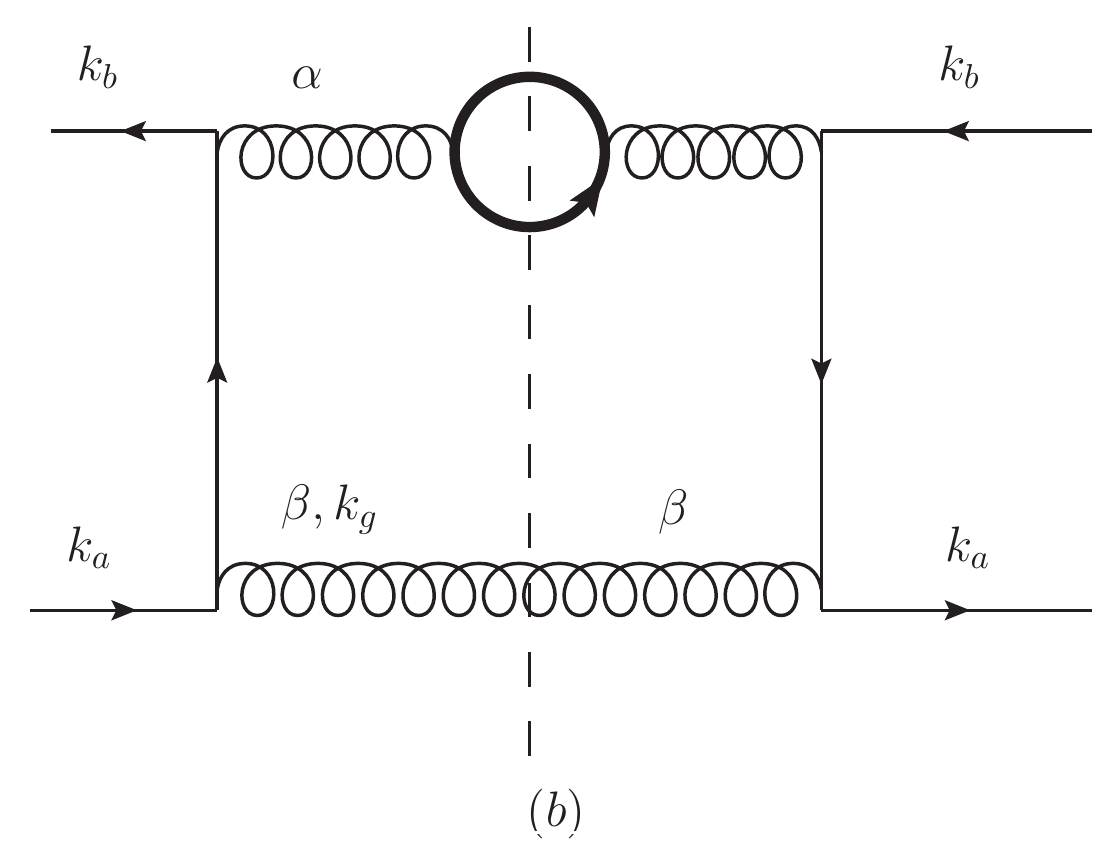}
\includegraphics[width=0.12\textwidth]{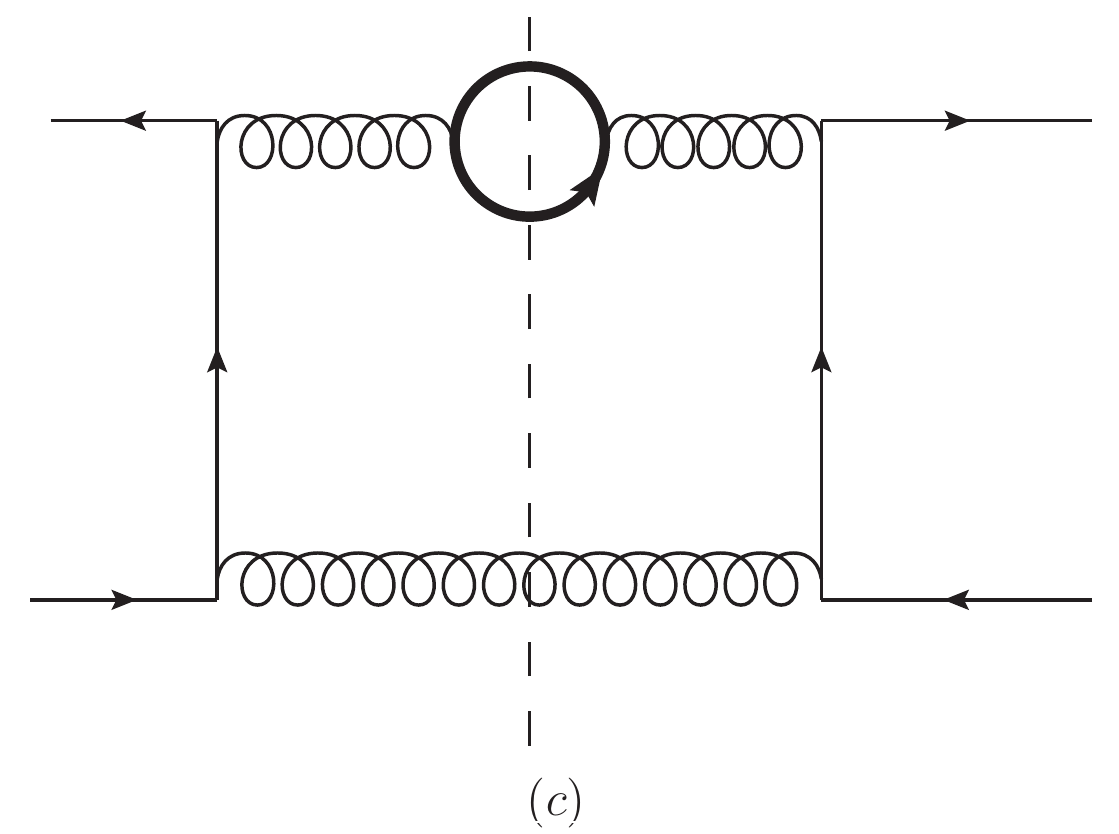}
\includegraphics[width=0.12\textwidth]{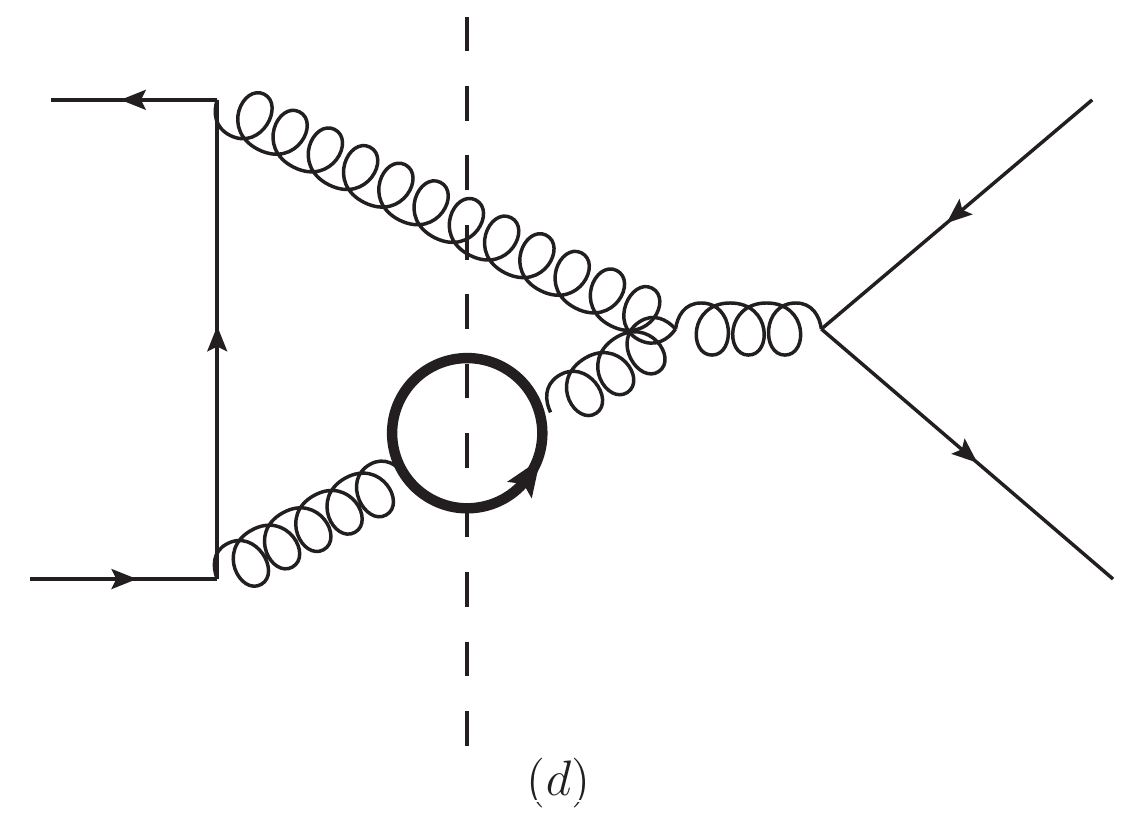}
\includegraphics[width=0.12\textwidth]{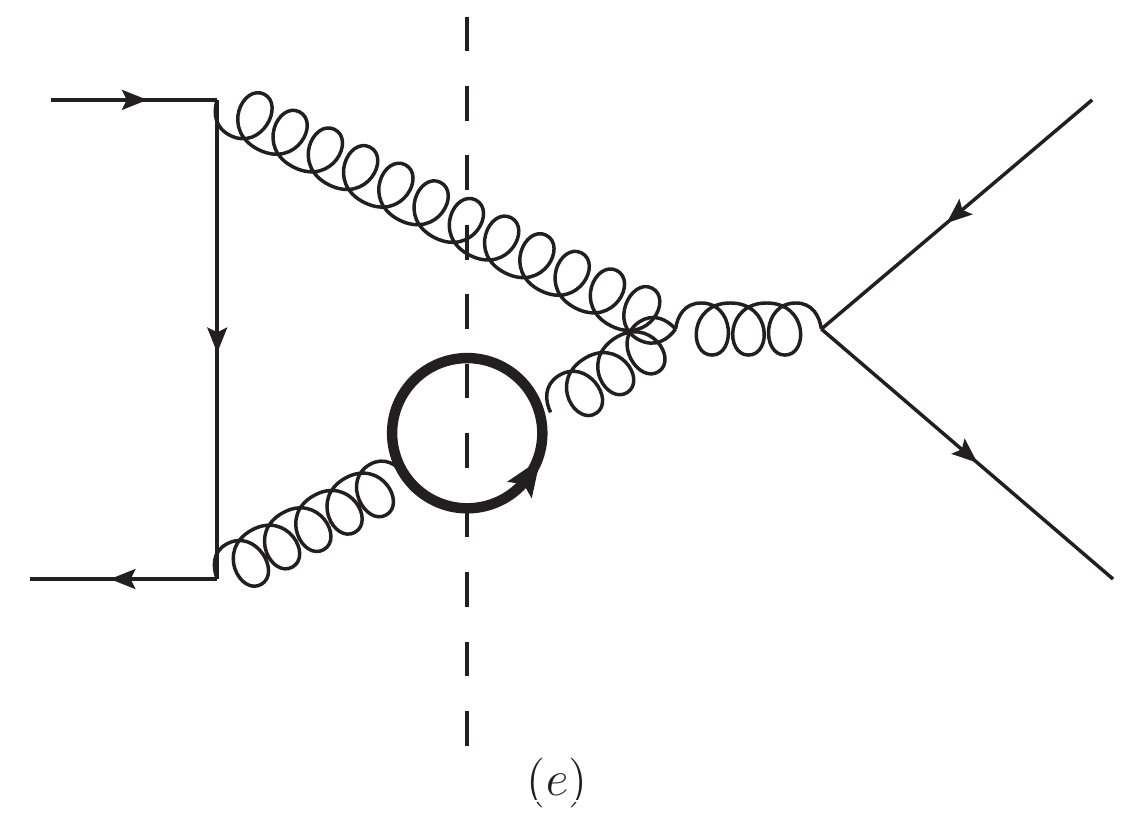}
\includegraphics[width=0.12\textwidth]{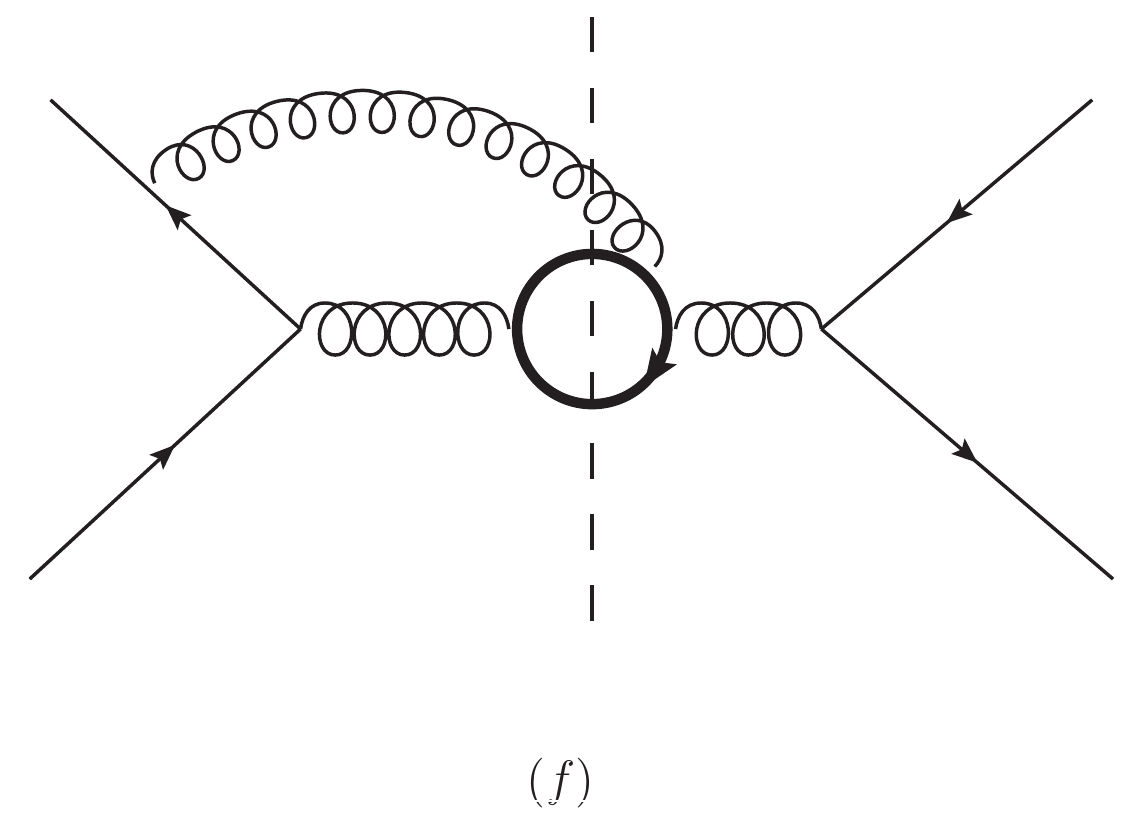}
\includegraphics[width=0.12\textwidth]{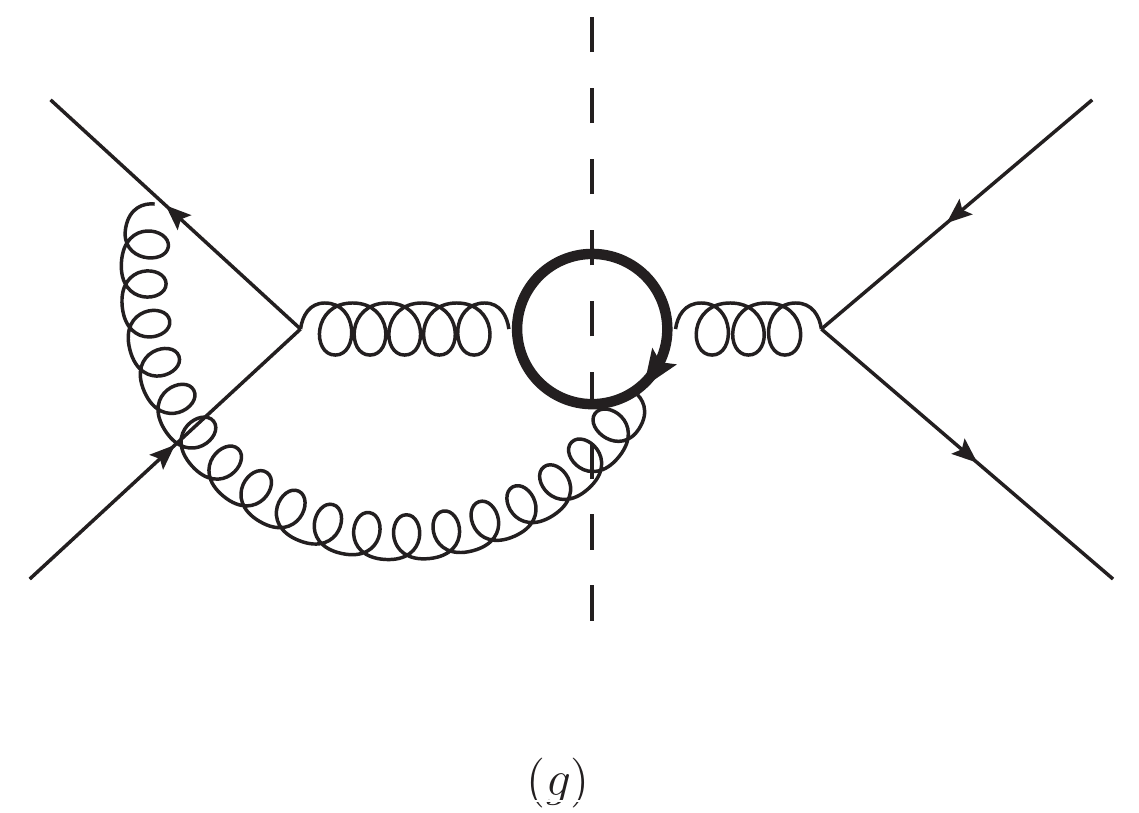}
\includegraphics[width=0.12\textwidth]{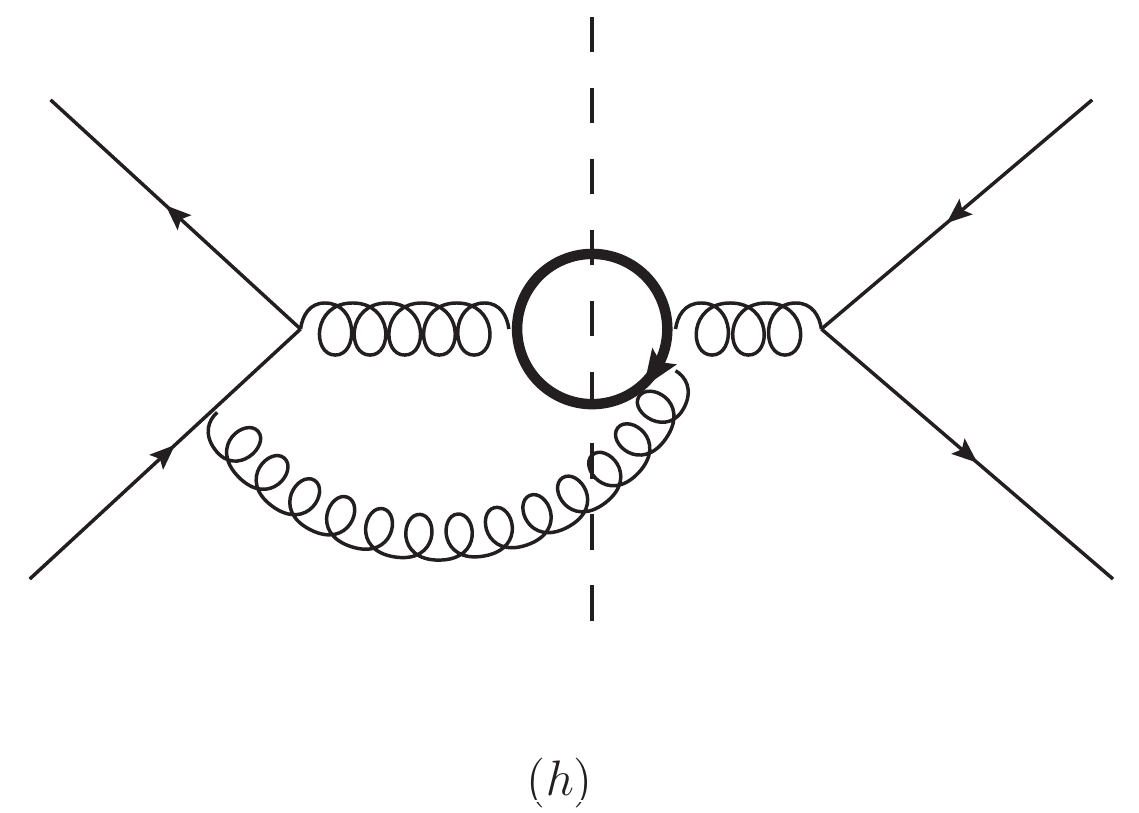}
\includegraphics[width=0.12\textwidth]{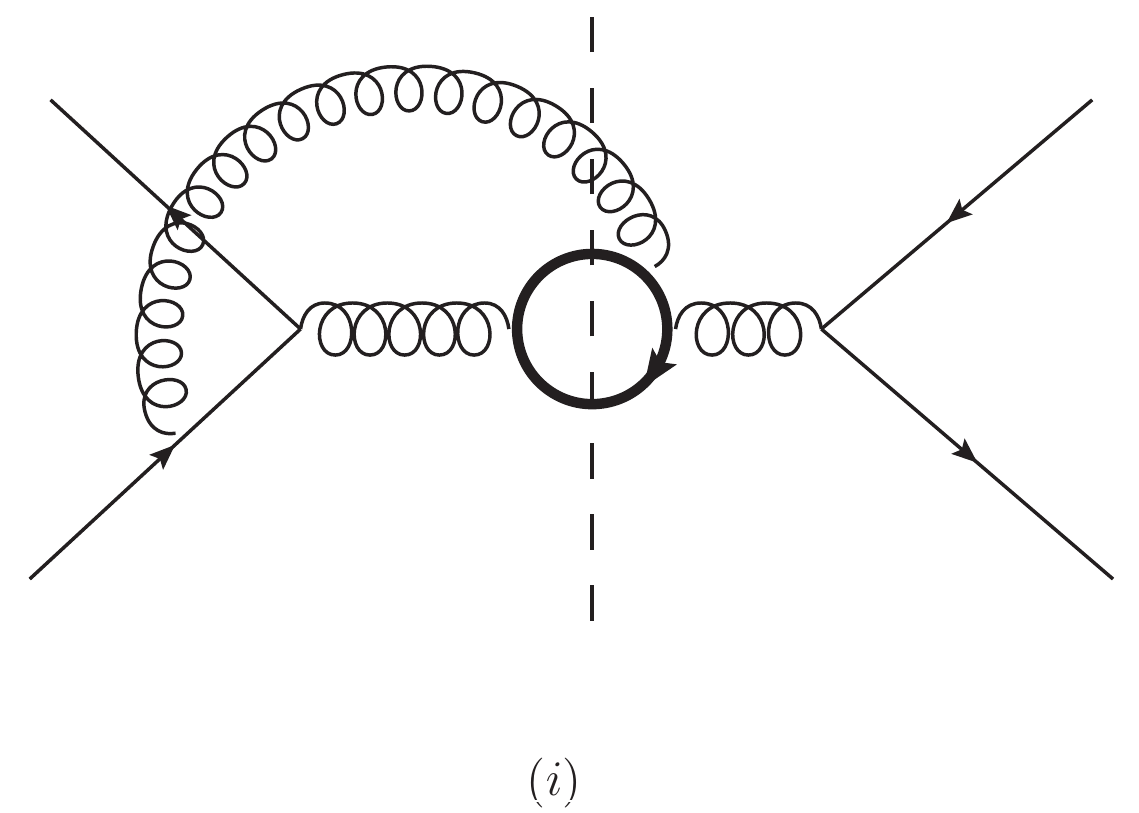}
\includegraphics[width=0.12\textwidth]{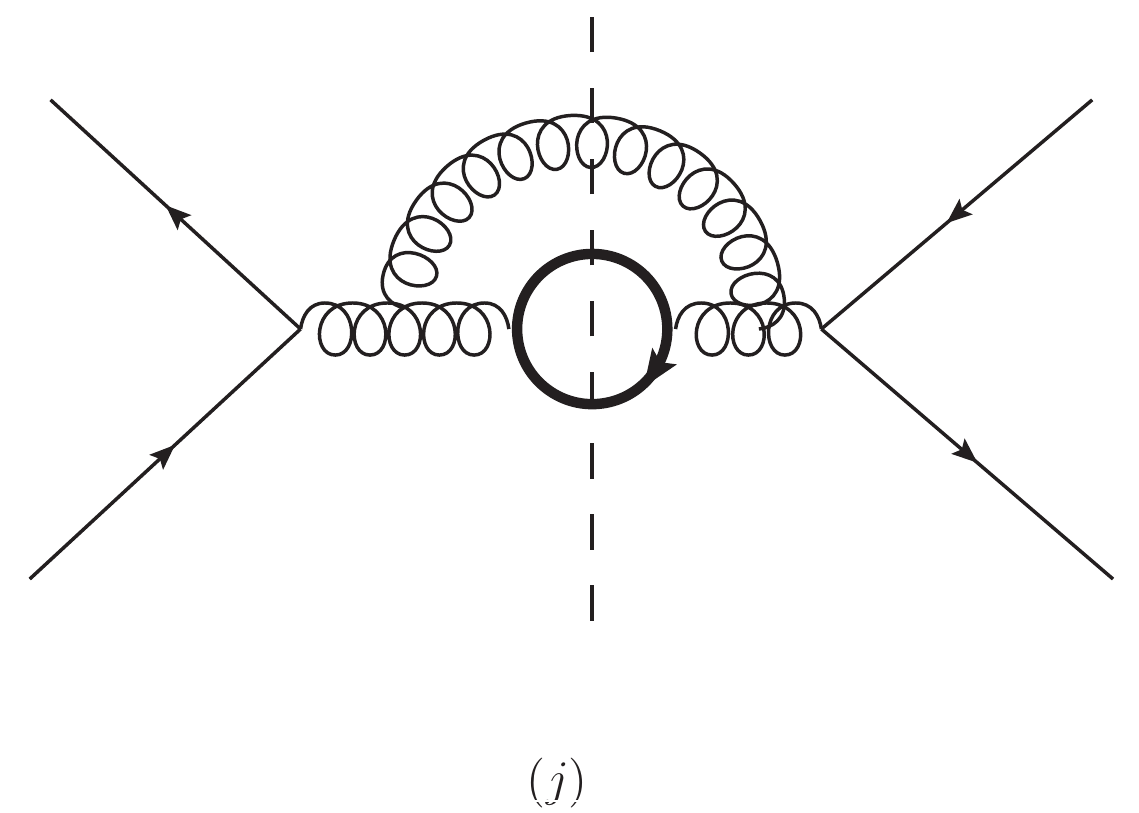}
\includegraphics[width=0.12\textwidth]{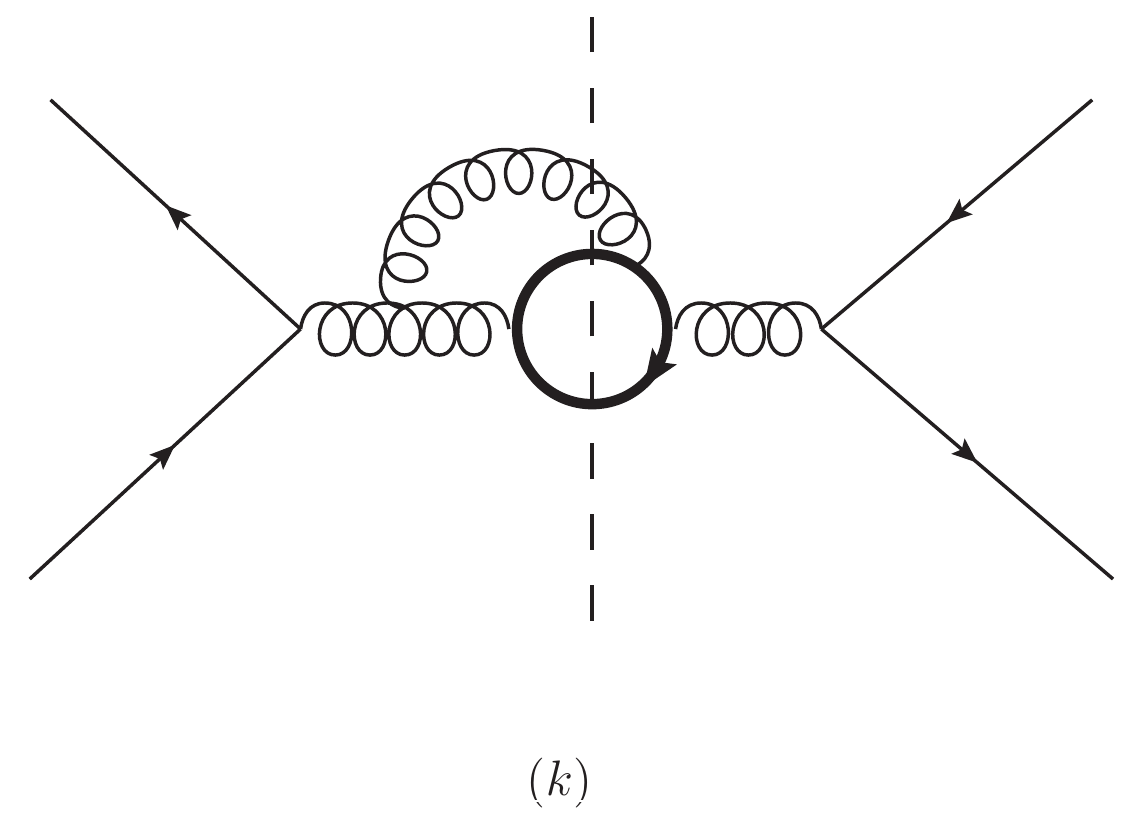}
\includegraphics[width=0.12\textwidth]{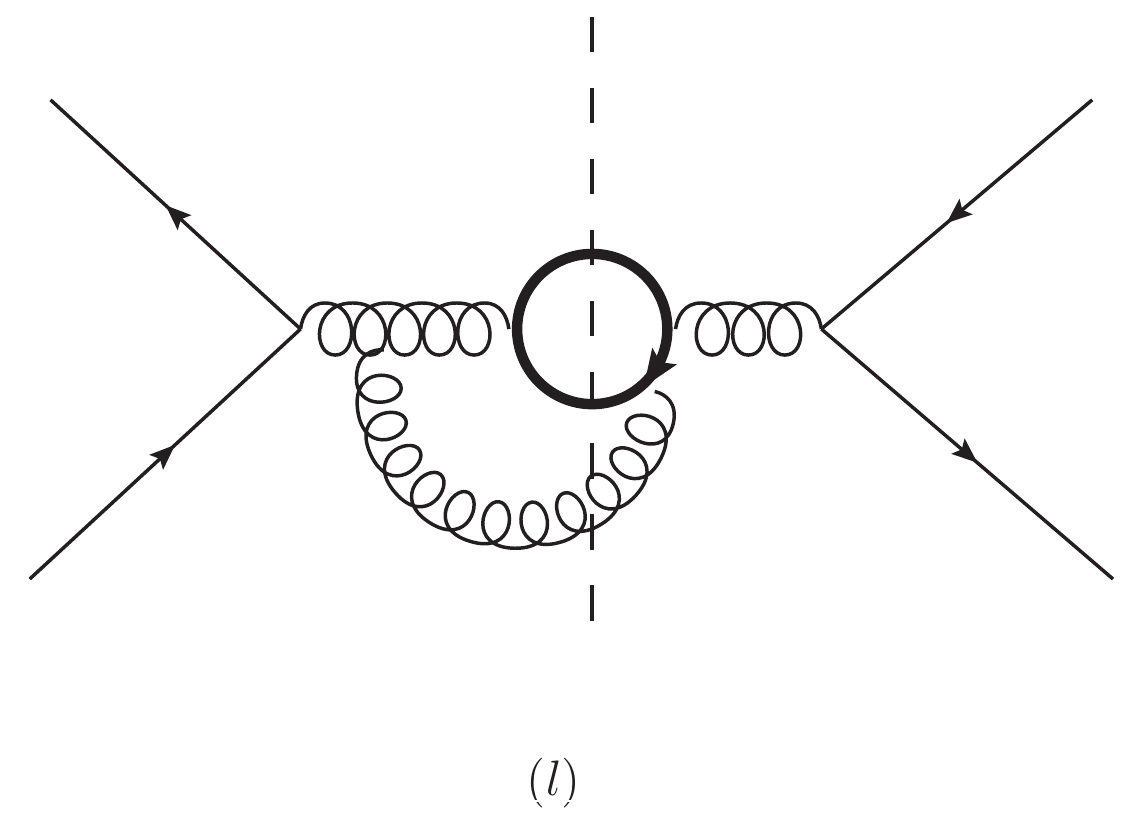}
\includegraphics[width=0.12\textwidth]{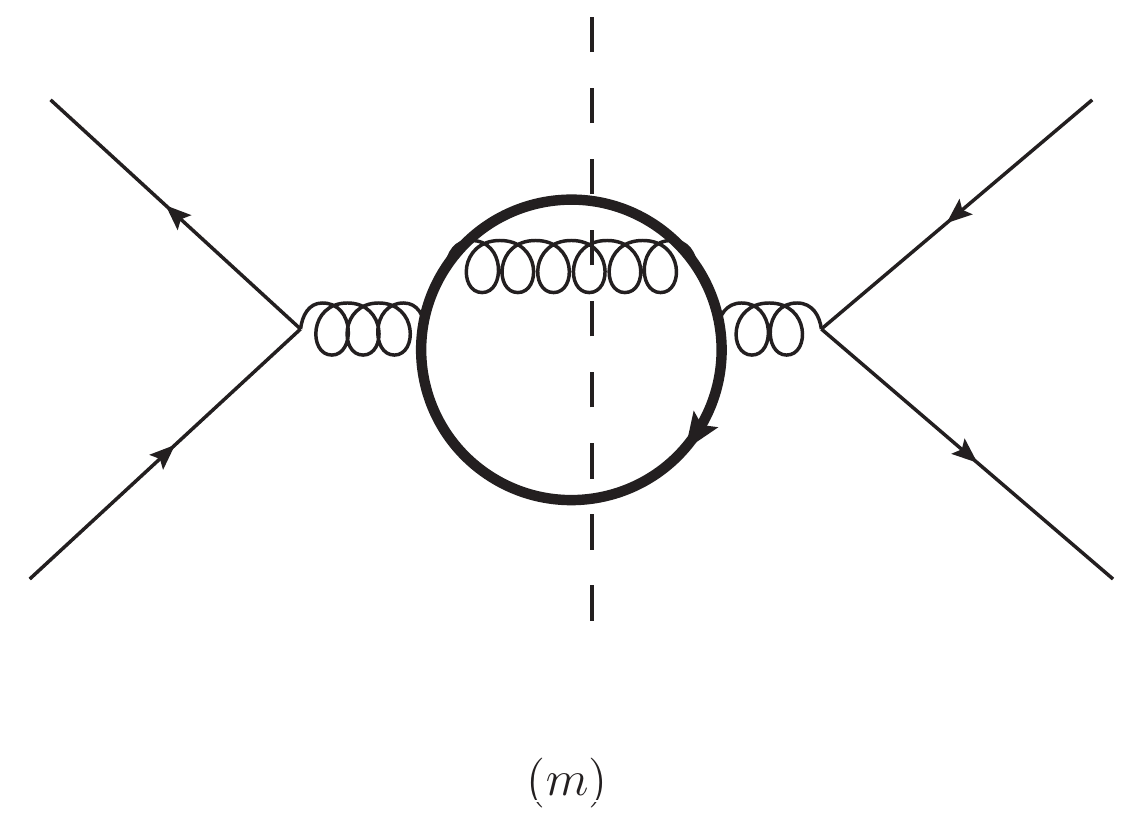}
\includegraphics[width=0.12\textwidth]{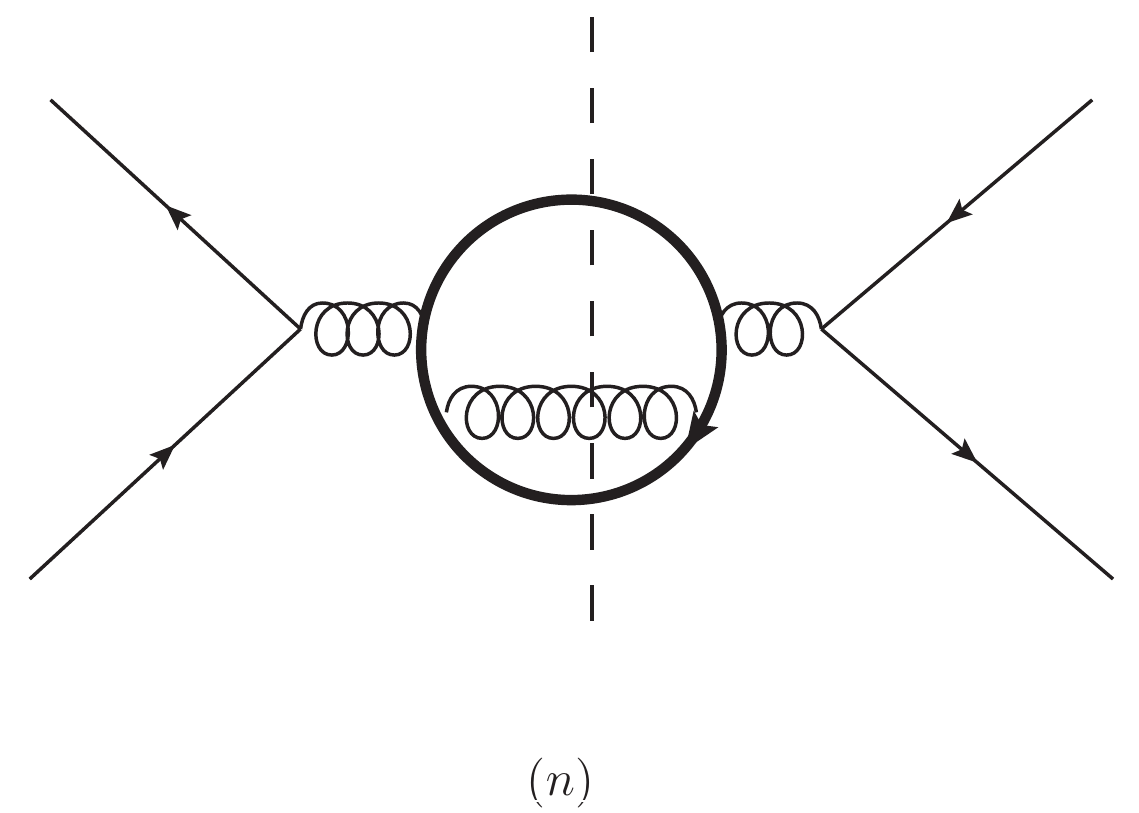}
\includegraphics[width=0.12\textwidth]{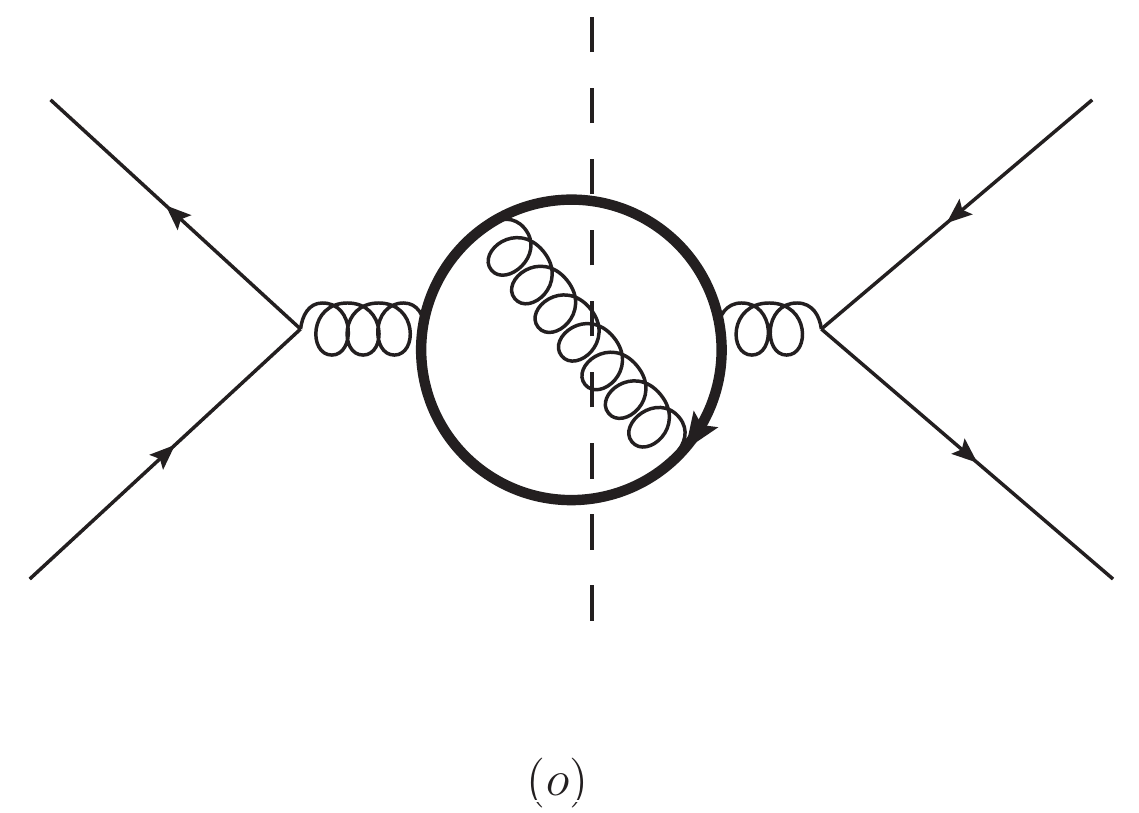}
\end{flushleft}
\caption{Diagrams for real corrections. Conjugated diagrams are not shown, but included in the calculation. }
\label{fig:real_correction}
\end{figure}
Real correction is given by following process,
\begin{align}
q(k_a)+\bar{q}(k_b)\rightarrow Q(p_1)+\bar{Q}(p_2)+g(k_g).
\end{align}
To get the cross section we calculate the cut diagrams in Fig.\ref{fig:real_correction}.
The partonic cross sections are obtained according to eq.(\ref{eq:fac_formula}).
Now the hard part $H_{ij}^{mn}$ in eq.(\ref{eq:fac_formula})
contains a two-body phase
space integration for $p_2$ and $k_g$. By moving heavy quark with
momentum $p_1$ from final state to initial state, the sum of these
cut diagrams is equal to the cut amplitude of following
forward scattering
\begin{align}
q(k_a)+\bar{q}(k_b)+\bar{Q}(-p_1)\rightarrow q(k_a)+\bar{q}(k_b)+\bar{Q}(-p_1),
\end{align}
with intermediate on-shell state being $|\bar{Q}(p_2),g(k_g)\rangle $.

\begin{figure}
\begin{center}
\includegraphics[width=0.15\textwidth]{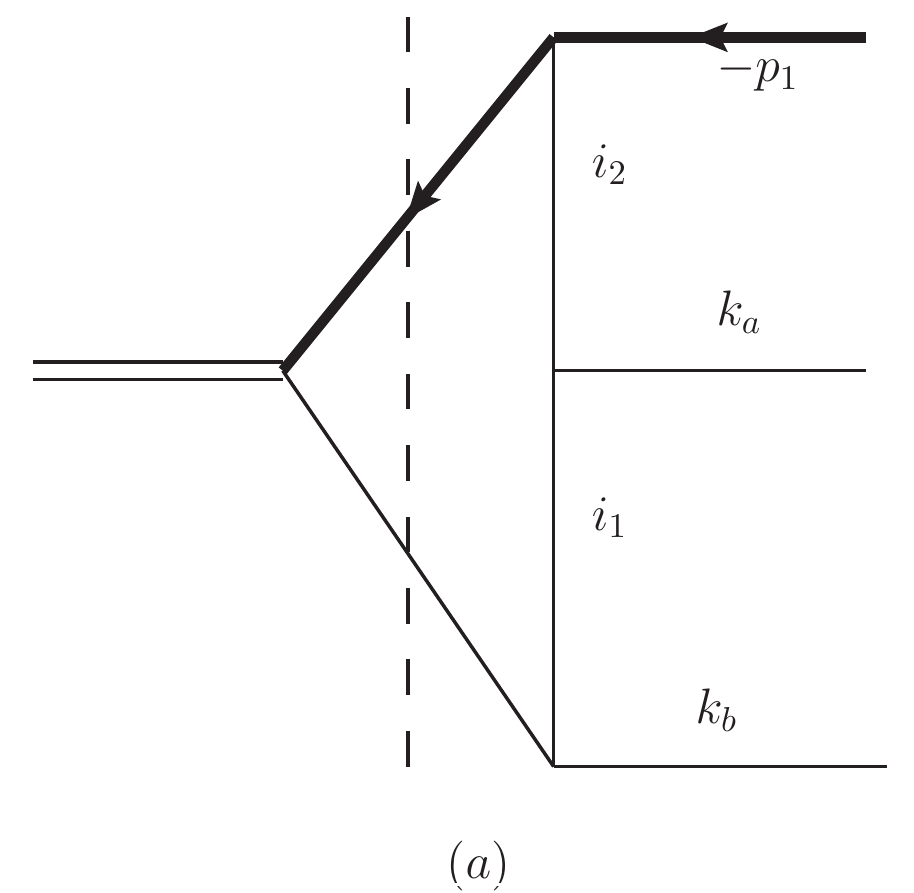}
\includegraphics[width=0.15\textwidth]{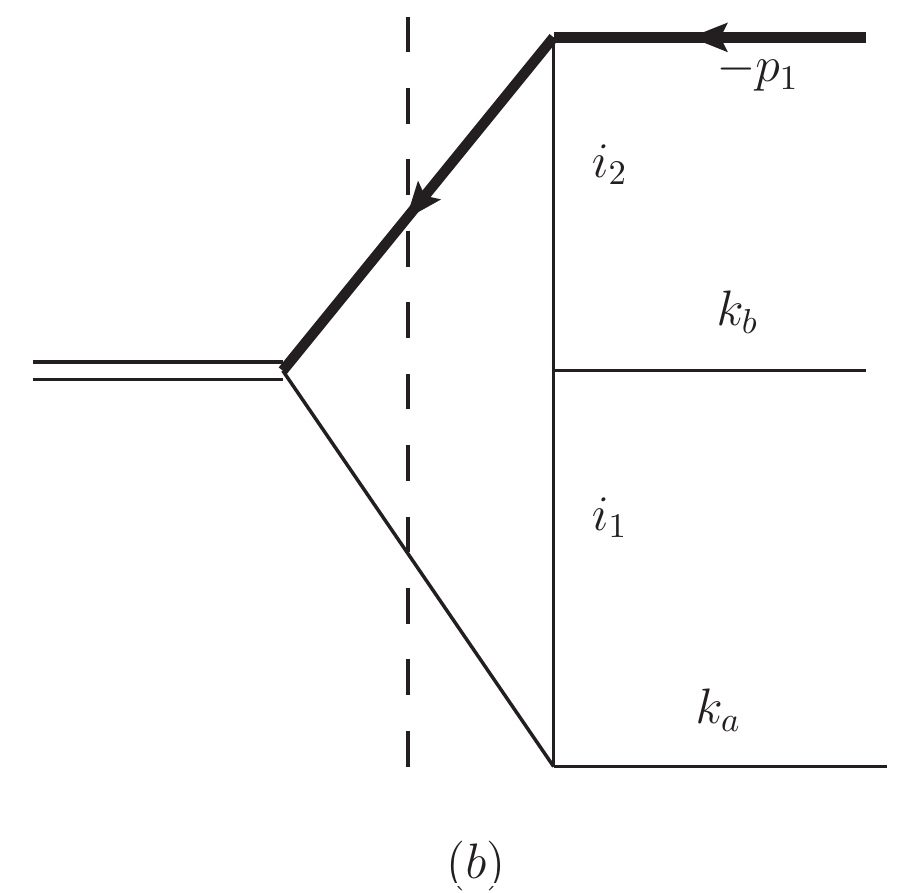}
\includegraphics[width=0.15\textwidth]{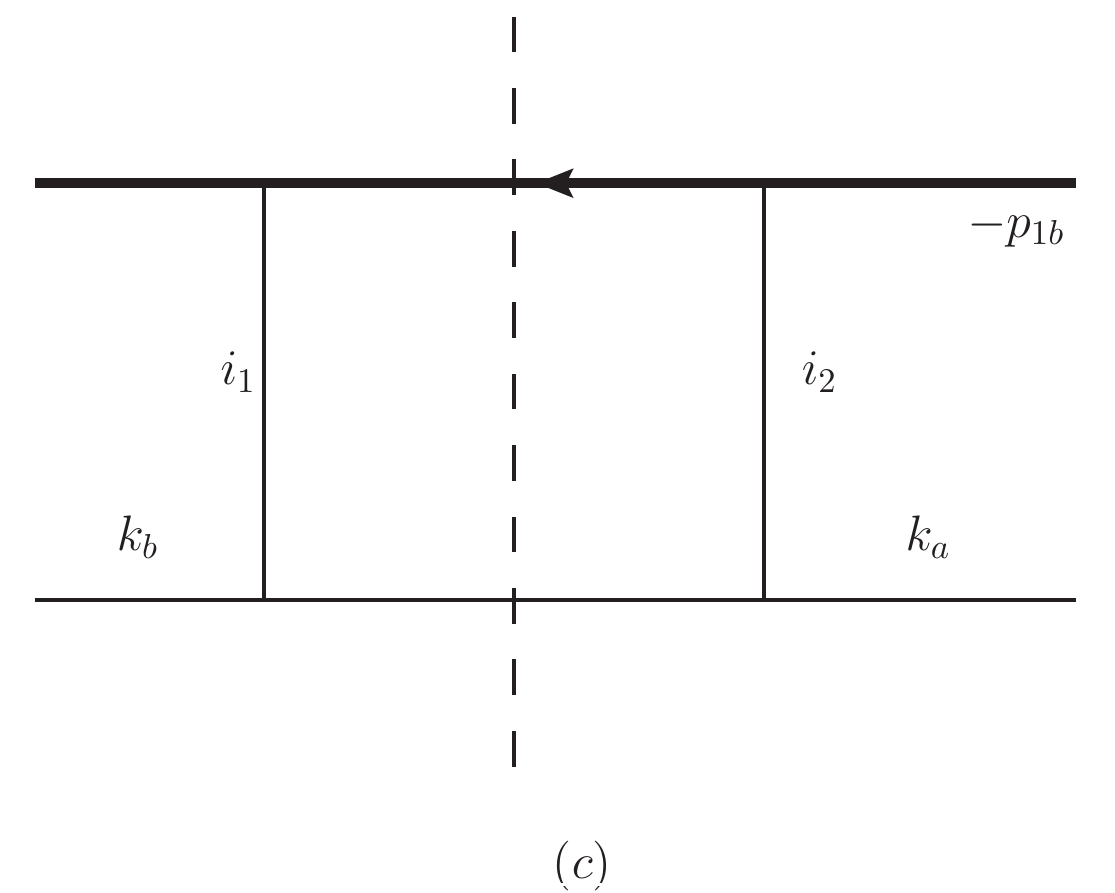}
\includegraphics[width=0.15\textwidth]{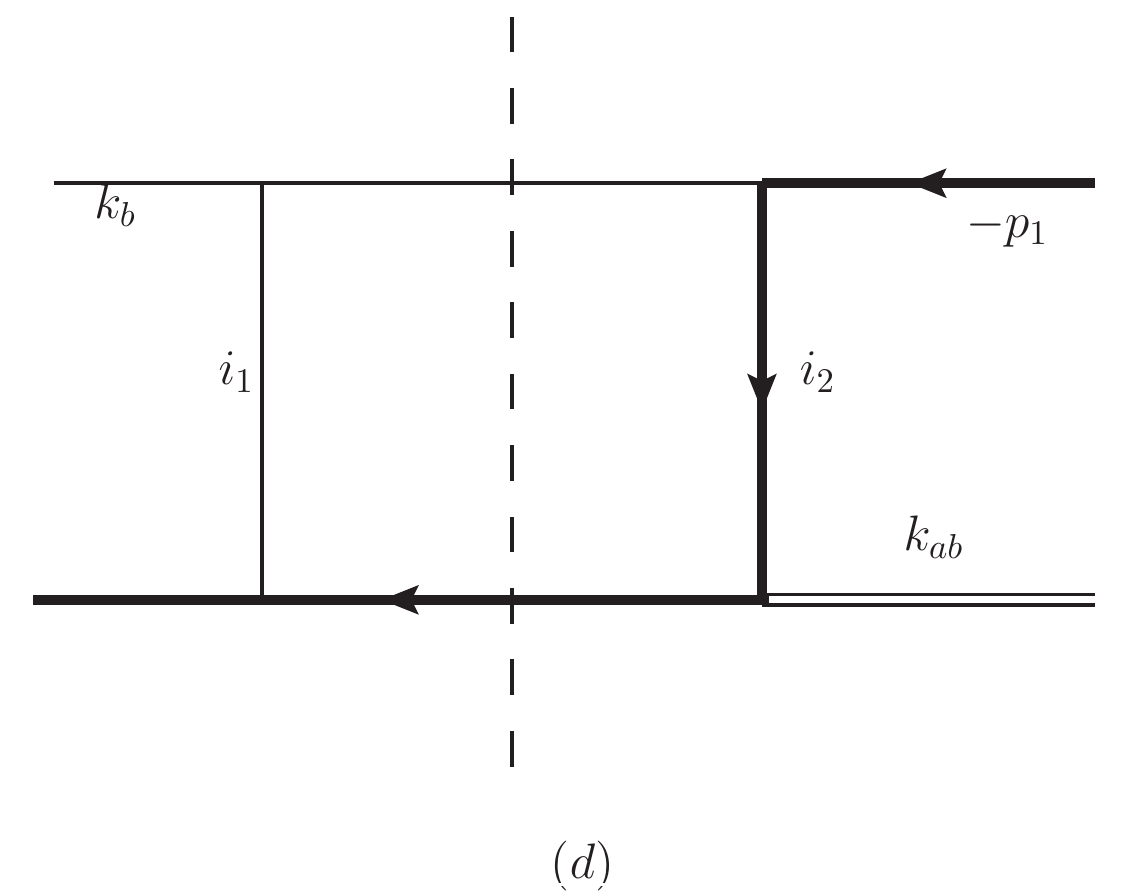}
\includegraphics[width=0.15\textwidth]{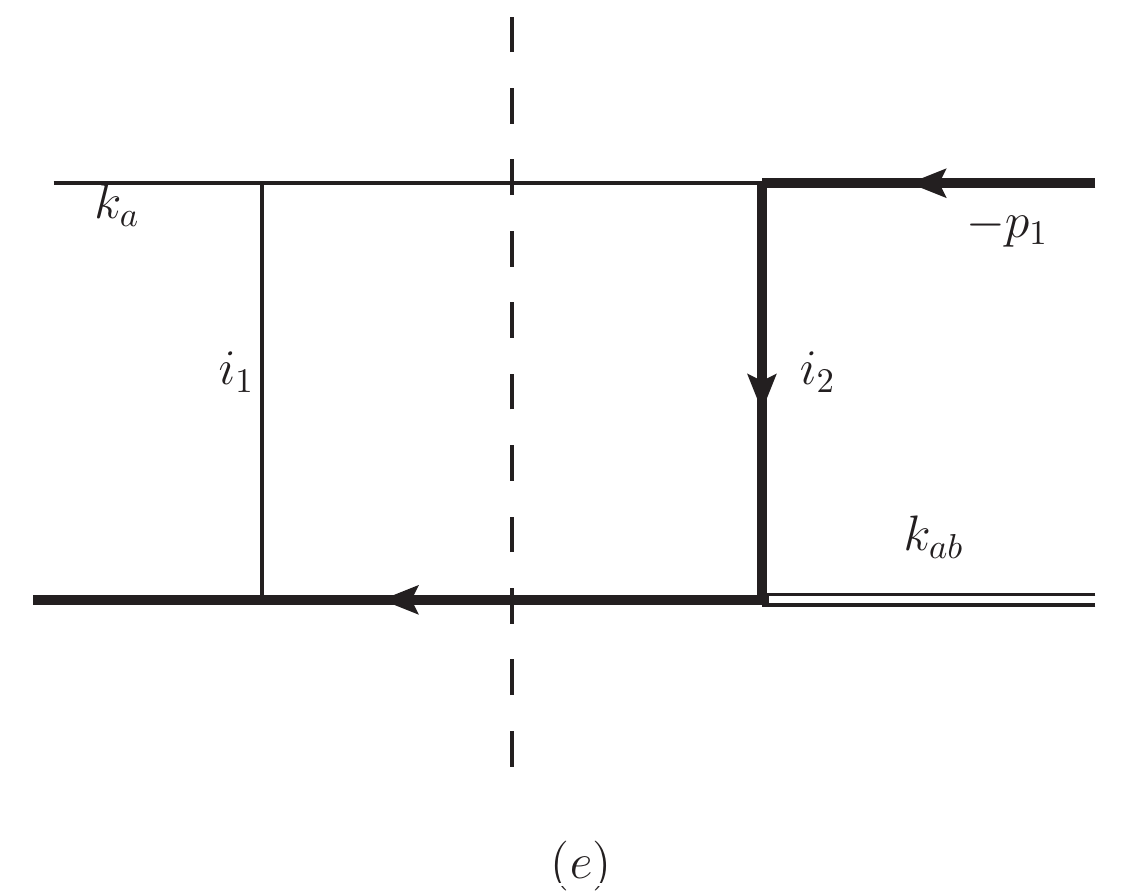}
\includegraphics[width=0.15\textwidth]{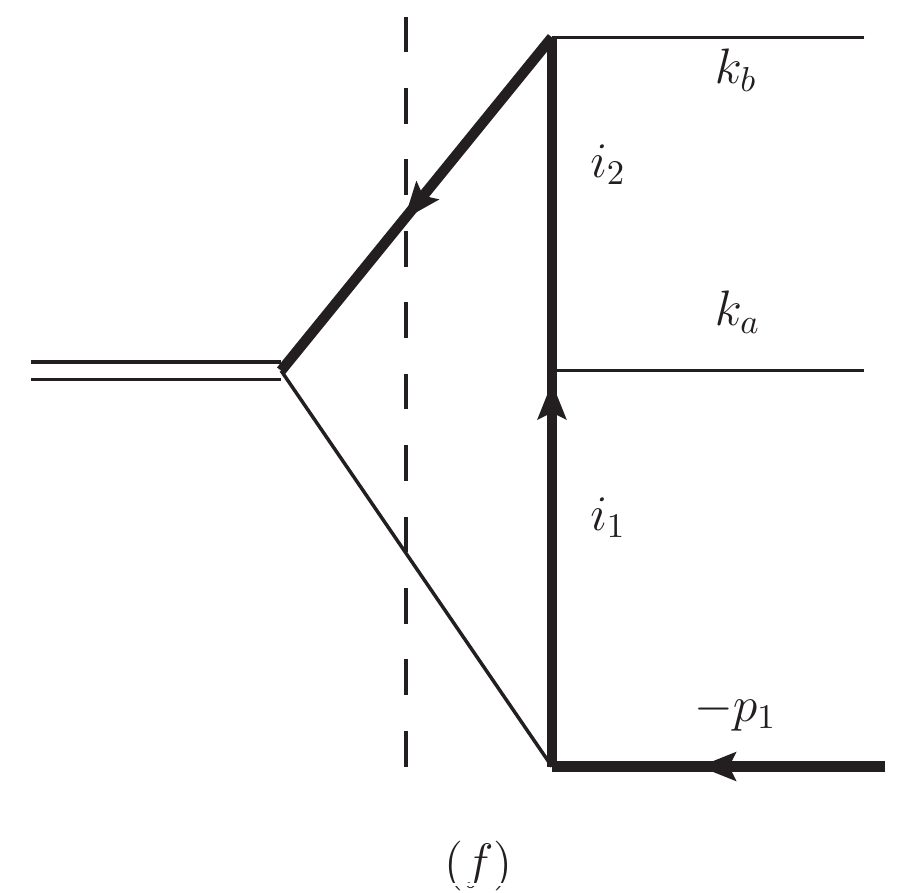}
\caption{Master integrals for real correction. $i_{1,2}=0,1$ are the indices of propagators. $k_{ab}=k_a+k_b$, $p_{1a}=p_1-k_a$, $p_{1b}=p_1-k_b$. External momenta on left-hand side of the cut are ingoing and those on right-hand side of the cut are outgoing. The thick line is for heavy quark. The thin real line is for massless parton. Off-shell external momenta are represented by double line. }
\label{fig:reduced_box}
\end{center}
\end{figure}

Then, the involved cut tensor integrals can be reduced to scalar ones in the same way as uncut tensor integrals\cite{Anastasiou:2002yz}. FIRE\cite{Smirnov:2008iw} with IBPs incorporated is a particularly suitable
tool for this purpose. After reduction, there are only six
types of master integrals, which are shown in Fig.\ref{fig:reduced_box}.
The general form of the master integral is
\begin{align}
I_r=\int \frac{d^n k_g}{(2\pi)^n}\frac{(2\pi)^2\de_1\de_2}{N_1^{i_1}N_2^{i_2}},
\hspace{5mm}
\de_1\equiv \de(k_g^2),\ \de_2\equiv\de((k_a+k_b-p_1-k_g)^2-m^2),
\end{align}
where $N_1,N_2$ are denominators of uncut propagators.
These $I_r$ integrals are in the standard form, i.e.,
$i_{1,2}=0$ or 1.

To calculate $I_r$ it is convenient to work in the frame with $\vec{q}=0$, where $q=k_a+k_b-p_1$. First, the energy of gluon $k_g^0$ can be integrated out by using the two delta functions. Then,
\begin{align}
I_r=& \lrb{\frac{\mu}{k_g^0}}^{\ep}\frac{k_g^0}{4q^0}
\int d\Omega_{n-1}\frac{1}{N_1^{i_1}N_2^{i_2}},
\end{align}
with $k_g^0$ the energy of final gluon. Explicitly,
$k_g^0=s\tau_x/(2q^0)$ and $q^0=\sqrt{m^2+s\tau_x}$. $d\Omega_{n-1}$
is the angular integration measure for $\vec{k}_g$, which is defined in $n-1=3-\ep$
dimensional space.  $I_r$ may contain
collinear and soft divergences. Different from virtual correction, these two divergences in real correction can be separated very easily.
Soft divergence corresponds to the singularity at $\tau_x=0$.
If $I_r$ is singular under soft limit $\tau_x \rightarrow 0$,
$N_1$ or $N_2$ must be proportional to $k_g^0$.
Then, $k_g^0$ can be extracted from $N_1$ or $N_2$. This implies we can define
an integral $\tilde{I}_r$ as follows, which is regular under soft limit.
\begin{align}
I_r=\lrb{\frac{\mu}{k_g^0}}^{\ep} \tau_x^k s^{-i_1-i_2}\tilde{I}_r.
\end{align}
$\tau_x^k$ is extracted from $N_1,N_2$. If $I_r$ contains soft divergence, $k=-1$.
In this way, collinear divergence is included in $\tilde{I}_r$ and soft divergence
is given by the expansion of $\tau_x^{-1-\ep}$ in $\ep$, i.e.,
\begin{align}
\tau_x^{-1-\ep}=\frac{1}{-\ep}\de(\tau_x)+\lrb{\frac{1}{\tau_x}}_+
-\ep\lrb{\frac{\ln\tau_x}{\tau_x}}_+.
\end{align}
The plus function is the standard one\cite{Nason:1989zy}.

The angular integrals $\tilde{I}_r$ can be classified into following six types.
\begin{align}
&R_1(w)=\int d\Omega_{n-1}\frac{1}{1+\vec{a}\cdot \vec{k}}
\frac{1}{1+\vec{b}\cdot \vec{k}},
&&R_2(\Delta,w)=\int d\Omega_{n-1}\frac{1}{1+\vec{a}\cdot \vec{k}}
\frac{1}{\Delta+\vec{b}\cdot \vec{k}},\no
&R_3(\delta,\Delta,w)=\int d\Omega_{n-1}\frac{1}{\de+\vec{a}\cdot \vec{k}}
\frac{1}{\Delta+\vec{b}\cdot \vec{k}},
&&R_4=\int d\Omega_{n-1}\frac{1}{1+\vec{a}\cdot \vec{k}},\no
&R_5(\de)=\int d\Omega_{n-1}\frac{1}{\de+\vec{a}\cdot \vec{k}},
&&R_6=\int d\Omega_{n-1}.
\label{eq:Ri}
\end{align}
$\vec{k},\vec{a},\vec{b}$ are normalized to one, i.e.,
$|\vec{a}|=|\vec{b}|=|\vec{k}|=1$. $w=1/|\vec{a}-\vec{b}|$ and
$\Delta>1,\de>1$. $d\Omega_{n-1}$ is for $\vec{k}$.

Same as the reduction of virtual integrals, IR pole $1/\ep_{IR}$ may transfer
from tensor integrals to reduced coefficients after FIRE reduction. Thus, some
$R_i$ should be expanded to higher orders of $\ep$. In our case, we have
checked that $R_6$ should be calculated to $O(\ep^2)$, while others, except for
$R_3$, should be expanded to $O(\ep)$. We just need $O(\ep^0)$ part of $R_3$.
By making use of Feynman parameters, these $R_i$ are calculated and the results are given in
Appendix.\ref{sec:Ri}. These results are compared with known results in \cite{Beenakker:1988bq}. Numerically, they are the same.

In real correction, $\Delta H_d$ and $H_d$ represent soft gluon contribution,
because $\tau_x=0$. These soft contributions can be obtained
by eikonal approximation and are factorized diagram by diagram.
Thus, we expect the real corrections to $\Delta H_d$ and $H_d$ are the same.
The explicit calculation confirms this. Interestingly, such soft correction for unpolarized $q\bar{q}$ scattering has been given in \cite{Beenakker:1990maa}.
For convenience, we show the result of \cite{Beenakker:1990maa} here.
\begin{align}
\{\Delta h_d^{(1)},h_d^{(1)}\}=& \{\Delta h_d^{(0)},h_d^{(0)}\}
\frac{1}{2}e^{-\frac{\ep}{2}(\ga_E-\ln 4\pi)}\lrb{\frac{s^2}{\mu^2 m^2}}^{-\ep/2}
\Big[ C_F K_{soft}^F+C_A K_{soft}^A \Big],\no
K_{soft}^F=&
\frac{16}{\ep^2}-\frac{8}{\ep}\ln{y}+2\ln^2y
+4Li_2\lrb{1-y}\no
&+4(1-\frac{2m^2}{s})\frac{1}{\be}\Big\{
\frac{2}{\ep}\ln x -\ln x +2Li_2(x)+2Li_2(-x)
-\ln^2 x+2\ln x\ln(1-x^2)-\zeta(2)
\Big\}\no
&+\frac{8}{\ep}+4-\frac{32}{\ep}\ln\frac{t_1}{u_1}-16\ln x \ln\frac{t_1}{u_1}
-16Li_2(1-\frac{u_1}{x t_1})
+16Li_2(1-\frac{t_1}{x u_1})
-6\zeta(2),\no
K_{soft}^A=&
\frac{4}{\ep}\ln y -\ln^2 y -2Li_2(1-y)\no
&-2(1-\frac{2m^2}{s})\frac{1}{\be}\Big\{
\frac{2}{\ep}\ln x+2Li_2(x)+2Li_2(-x)-\ln^2{x}+2\ln{x}\ln(1-x^2)
-\zeta(2)
\Big\}\no
&-\frac{12}{\ep}\ln\frac{u_1}{t_1}+6\ln{x}\ln\frac{t_1}{u_1}-\ln^2{x}
+\ln^2\frac{t_1}{u_1}
-6Li_2(1-\frac{t_1}{x u_1})+6Li_2(1-\frac{u_1}{x t_1}),
\label{eq:real_hd}
\end{align}
with
\begin{align}
y=\frac{s m^2}{t_1 u_1},\ x=\frac{1-\be}{1+\be},\ \be=\sqrt{1-\frac{4m^2}{s}},\
t_1=(k_a-p_1)^2-m^2=-s\tau_1,\ u_1=(k_b-p_1)^2-m^2=-s\tau_2.
\end{align}
We have checked that this result, including the finite part, is the same as
our result. This is a strong check for our reduction scheme and the calculation of
real integrals.

In hard coefficients $h_p$ and $h_l$, $\tau_x$ can be nonzero. Thus, $h_p$ and
$\Delta h_p$ contain collinear divergence only. $h_l, \Delta h_l$ are finite
and given by $-\ep h_p,-\ep \Delta h_p$, respectively. The collinear divergence
for transversely polarized $q\bar{q}$ scattering is
\begin{align}
\Delta h_p^{(1)}=\Delta h_d^{(0)}
\frac{-32\pi C_F R_\ep}{\ep \tau_1\tau_2(\rho-4\tau_1(1-\tau_1))}\lrb{\frac{\mu^2}{m^2}}^{\ep/2}
(\rho-4\tau_1\tau_2)(2\tau_1\tau_2+\tau_x(1-\tau_x)).
\label{eq:real_ts_hp}
\end{align}
Note that because $\Delta h_d^{(0)}\propto (\rho-4\tau_1(1-\tau_1))$,
$\Delta h_p^{(1)}$ is symmetric in $\tau_1,\tau_2$.

But the divergence of unpolarized $h_p^{(1)}$ is much more complicated,
\begin{align}
h_p^{(1)}=&h_d^{(0)}\frac{-16\pi C_F R_\ep}{\ep[2+\rho-4\tau_1(1-\tau_1)]}
\lrb{\frac{\mu^2}{m^2}}^{\ep/2}\Big[
\rho
\frac{-2(\tau_1^3+\tau_2^3)+(\tau_1^2+\tau_2^2)(1+\tau_1^2+\tau_2^2)}{
\tau_1^2\tau_2^2}\no
&+
\frac{r^6-r^4(1-\tau_x^2)-r^2(1+2\tau_x^2+8\tau_x^3+\tau_x^4)
+(1-\tau_x^2)(1+\tau_x^2)^2}{4\tau_1\tau_2(1-\tau_1)(1-\tau_2)}
\Big],
\label{eq:real_unp_hp}
\end{align}
with $r=\tau_1-\tau_2$.
To understand the difference between $h_p^{(1)}$ and $\Delta h_p^{(1)}$ is
interesting. First, $\Delta h_p$ and $h_p$ contain
only collinear divergence. Except for the ladder diagrams Fig.\ref{fig:real_correction}(a,b),
all other diagrams generate collinear divergence just from longitudinal real gluon. For longitudinal gluon,
Ward Identities can be applied and the summed result
can be expressed as the convolution of tree level partonic cross section and
gauge link correction to PDFs(see \cite{Collins:2008sg} for example). The latter is the same for unpolarized PDF and
transversity PDF. However, the divergence from ladder diagram is different.
Consider ladder diagram Fig.\ref{fig:real_correction}(b). For unpolarized case, according to the formula eq.(\ref{eq:fac_formula}),
the contribution of this diagram is proportional to
\begin{align}
I=\int d^n k_g \ga^+\ga^\al\frac{\s{k_a}-\s{k}_g}{(k_a-k_g)^2}\ga^\be\ga^-
\ga_\be\frac{\s{k}_a-\s{k}_g}{(k_a-k_g)^2}\de(k_g^2),
\end{align}
where $\ga^+,\ga^-$ are the projection matrices for unpolarized PDF.
Due to $\ga^-$, the
real gluon must be transverse, that is, $\ga^\be=\ga_\perp^\be$. Then, in collinear
region, it is $k_{g\perp}$ in the numerator that gives leading power contribution.
Because $\ga_\perp^\be\ga^-=-\ga^-\ga_\perp^\be$, the integral becomes
\begin{align}
I=\int d^n k_g \Big[\ga^+\ga^\al\ga^-\Big]
\frac{(2-\ep)k_{g\perp}^2}{[(k_a-k_g)^2]^2}\de(k_g^2).
\end{align}
It is clear that the part in brackets gives tree level result and $I$ contains a collinear divergence. On the other hand, for transversity contribution, this diagram is proportional to
\begin{align}
\Delta I=\int d^n k_g (\ga_5\s{s}_{b\perp}\ga^+)
\ga^\al\frac{\s{k_a}-\s{k}_g}{(k_a-k_g)^2}\ga^\be(\ga_5\s{s}_{a\perp}\ga^-)
\ga_\be\frac{\s{k}_a-\s{k}_g}{(k_a-k_g)^2}\de(k_g^2).
\end{align}
Due to the same reason, in collinear region $\Delta I$ receives contribution only from transverse gluon and can be written as
\begin{align}
\Delta I=\int d^n k_g (\ga_5\s{s}_{b\perp}\ga^+)
\ga^\al(-\ga_5\ga^-)
\frac{\s{k}_{g\perp}}{(k_a-k_g)^2}\ga_\perp^\be(\s{s}_{a\perp})
\ga_{\perp\be}\frac{\s{k}_{g\perp}}{(k_a-k_g)^2}\de(k_g^2).
\end{align}
By power counting in collinear region, $k_{g\perp}$ in the other part of this diagram can be ignored, so, in the integrand the replacement $k_{g\perp}^\mu k_{g\perp}^\nu\rightarrow k_{g\perp}^2 g_\perp^{\mu\nu}/(2-\ep)$ is allowed
and we get
\begin{align}
\Delta I=\int d^n k_g (\ga_5\s{s}_{b\perp}\ga^+)
\ga^\al(-\ga_5\ga^-)
\Big[\ga_\perp^\rho\ga_\perp^\be \s{s}_{a\perp}\ga_{\perp\be}
\ga_{\perp\rho}\Big]\frac{k_{g\perp}^2/(2-\ep)}{[(k_a-k_g)^2]^2}
\de(k_g^2).
\end{align}
Note that $\ga_\perp^\be\ga_\perp^\mu\ga_{\perp\be}=\ep \ga_\perp^\mu$. The
quantity in brackets becomes
\begin{align}
[\ga_\perp^\rho\ga_\perp^\be \s{s}_{a\perp}\ga_{\perp\be}
\ga_{\perp\rho}]=\ep^2 \s{s}_{a\perp}.
\end{align}
It is proportional to $\ep^2$! Thus the integral $\Delta I$ and then the ladder
diagram Fig.\ref{fig:real_correction}(b) just vanishes in collinear region when the limit $\ep\rightarrow 0$ is taken. This explains why $\Delta h_p^{(1)}$ is much
simpler than $h_p^{(1)}$.

\subsection{Subtraction and Final result}
To get the true one-loop contribution, we have to subtract collinear contributions
from each diagram\cite{Collins:2008sg}. The subtraction is realized by following replacement
in tree level hadron cross sections,
\begin{align}
\{f_1(x_a,\mu^2),h_1(x_a,\mu^2)\}\rightarrow & \frac{\al_s}{2\pi}
\frac{(4\pi)^{\ep/2}}{\Gamma(1-\ep/2)}
\Big[\frac{2}{\ep_{UV}}-\frac{2}{\ep_{IR}}\Big]
\int_{x_a}^1 \frac{d\xi_a}{\xi_a}
\{P_{qq}(\frac{x_a}{\xi_a})f_1(\xi_a,\mu^2),
P^T_{qq}(\frac{x_a}{\xi_a})h_1(\xi_a,\mu^2)
\}.
\end{align}
The DGLAP evolution kernels( see \cite{Brock:1993sz,Mukherjee:2001zx,Blumlein:2001ca} and reference therein) are
\begin{align}
P_{qq}(x)=C_F\Big[\frac{3}{2}\de(1-x)+\frac{1+x^2}{(1-x)_+}\Big],\
P^T_{qq}(x)=C_F\Big[\frac{3}{2}\de(1-x)+\frac{2x}{(1-x)_+}\Big].
\end{align}
The UV pole $2/\ep_{UV}$ is removed by renormalization(in $\msb$-scheme)
of bare transversity distribution which appearing in tree level cross section.
Then only IR pole should be preserved. The final subtraction terms are
\begin{align}
\frac{d\sig^{unp}}{dy d^2p_{1\perp}}\Big|_{sub}
=&\frac{\al_s C_F}{2\pi}
\frac{(4\pi)^{\ep/2}}{\Gamma(1-\frac{\ep}{2})}
\lrb{\frac{-2}{\ep_{IR}}}
\int dx_a dx_b  f_1(x_a,\mu^2)\bar{f}_1(x_b,\mu^2)
\Big[\frac{z_a P_{qq}(z_a)}{1-\tau_1}H_d(z_a x_a,x_b)
+\frac{z_b P_{qq}(z_b)}{1-\tau_2}H_d(x_a,z_b x_b)
\Big]\no
=&\frac{\al_s C_F}{2\pi}
\frac{(4\pi)^{\ep/2}}{\Gamma(1-\frac{\ep}{2})}
\lrb{\frac{-2}{\ep_{IR}}}
\int dx_a dx_b  f_1(x_a,\mu^2)\bar{f}_1(x_b,\mu^2)\times\no
&\Big[
(3-2\ln\tau_1(1-\tau_1))\de(\tau_x)H_d(x_a,x_b)
+\frac{2}{(\tau_x)_+}\Big(z_a(1+z_a^2) H_d(z_a x_a, x_b)
+z_b(1+z_b^2) H_d(x_a,z_b x_b)\Big)
\Big],\no
\frac{d\Delta\sig}{dy d^2p_{1\perp}}\Big|_{sub}
=&\cos(2\phi)|s_{a\perp}||s_{b\perp}|
\frac{\al_s C_F}{2\pi}
\frac{(4\pi)^{\ep/2}}{\Gamma(1-\frac{\ep}{2})}
\lrb{\frac{-2}{\ep_{IR}}}\times\no
&\int dx_a dx_b  h_1(x_a,\mu^2)\bar{h}_1(x_b,\mu^2)
\Big[\frac{z_a P^T_{qq}(z_a)}{1-\tau_1}\Delta H_d(z_a x_a,x_b)
+\frac{z_b P^T_{qq}(z_b)}{1-\tau_2}\Delta H_d(x_a,z_b x_b)
\Big]\no
=&\cos(2\phi)|s_{a\perp}||s_{b\perp}|
\frac{\al_s C_F}{2\pi}
\frac{(4\pi)^{\ep/2}}{\Gamma(1-\frac{\ep}{2})}
\lrb{\frac{-2}{\ep_{IR}}}
\int dx_a dx_b  h_1(x_a,\mu^2)\bar{h}_1(x_b,\mu^2)
\times\no
&\Big[(3-2\ln\tau_1(1-\tau_1))\de(\tau_x)\Delta H_d(x_a,x_b)
+\frac{2}{(\tau_x)_+}\Big(z_a^2 \Delta H_d(z_a x_a, x_b)
+z_b^2 \Delta H_d(x_a,z_b x_b)\Big)
\Big],
\label{eq:sub_terms}
\end{align}
with $z_a=\tau_2/(1-\tau_1),z_b=\tau_1/(1-\tau_2)$.
The logarithm before $\de(\tau_x)$ comes from the variable transformation
in plus function\cite{Nason:1989zy},
\begin{align}
\lrb{\frac{1}{a\tau_x}}_+=&\frac{1}{a}\lrb{\frac{1}{\tau_x}}_++\frac{\ln a}{a}\de(\tau_x).
\end{align}
Note that the subtraction terms contain no explicit $\ln\mu$. The final one-loop
(order $\al_s^3$) cross sections are given by
\begin{align}
\frac{d\sig}{dy d^2p_{1\perp}}=&
\frac{d\sig}{dy d^2p_{1\perp}}\Big|_{real+vir} - \frac{d\sig}{dy d^2p_{1\perp}}\Big|_{sub}.
\end{align}
The subtraction term is given by eq.(\ref{eq:sub_terms}),
and the divergent part of real and
virtual corrections are given by eqs.(\ref{eq:virtual_correction},\ref{eq:real_hd},
\ref{eq:real_ts_hp},\ref{eq:real_unp_hp}).
From these results, the soft divergences
appearing in real and virtual corrections are cancelled, and the remaining collinear
divergences are removed by the subtraction terms. The final one-loop cross section
is then finite. Further, we note that massless wave function renormalization, UV
counter terms and subtraction terms all are $\ln\mu$ independent. The $\ln\mu$
dependence comes from the loop integrals in Figs.\ref{fig:virtual},\ref{fig:real_correction}, and from the massive
fermion wave function renormalization constant $Z_2^{(m)}$ in eq.(\ref{eq:wave_function}).
The $\ln\mu$ dependence extracted from above results is
\begin{align}
\frac{d\Delta \sig}{dy d^2 p_{1\perp}}=&
\int dx_a dx_b h_1(x_a,\mu^2)
\bar{h}_1(x_b,\mu^2)\frac{\al_s^2}{s}\Delta h_d^{(0)}(x_a,x_b)\de(\tau_x)\no
&+
\int dx_a dx_b h_1(x_a,\mu^2)
\bar{h}_1(x_b,\mu^2)\ln\frac{\mu^2}{m^2}
\Big\{
\al_s\times (2b_0) \Delta H_d^{tree}(x_a,x_b)\de(\tau_x)\no
&-\frac{\al_s C_F}{2\pi}\Big[
\frac{z_a}{1-\tau_1}P^T_{qq}(z_a)\Delta H_d^{tree}(z_a x_a,x_b)
+\frac{z_b}{1-\tau_2}P^T_{qq}(z_b)\Delta H_d^{tree}(x_a,z_b x_b)
\Big]
\Big\}+\cdots,
\label{eq:mu-dependence}
\end{align}
where $\cdots$ are of order $\al_s^3$ and do not depend on $\ln\mu$ explicitly.
Note that $\Delta H_d^{tree}=(\al_s^2/s) \Delta h_d^{(0)}$ and the $\mu$ dependence of $\al_s$ is given by RGE
\begin{align}
\frac{\partial\al_s(\mu^2)}{\partial \ln\mu^2}=-b_0\al_s^2(\mu^2),\
b_0=\frac{11C_A-2(2+n_F)}{12\pi}.
\end{align}
Now it is obvious that eq.(\ref{eq:mu-dependence}) is $\ln\mu$ independent up to $O(\al_s^4)$.
For unpolarized cross section, the same
conclusion holds by transparent replacement of PDFs and DGLAP evolution kernels.
Now, we finish our calculation of one-loop correction to heavy quark production
process. All results, including finite hard coefficients, are stored in
mathematica files which can be obtained from author if required.

Before ending this section, we want to discuss the regularization of $\ga_5$ in
dimensional scheme. This problem, however, is related to the regularization
of spin vector $s^\mu$. For a consistent regularization scheme, if the momentum
of a particle is defined in n dimensional space, the related spin vector or
polarization vector should also be defined in n dimensional space. For our
case here, the hadron momenta $P_A,P_B$ are external momenta, which are allowed
to be constrained in 4 dimensional space. Thus, spin vectors $s_{a}^\mu,s_{b}^\mu$
are defined in 4 dimensional space. Then, we consider HVBM scheme for $\ga_5$
\cite{tHooft:1972tcz,Breitenlohner:1977hr}. In
this scheme $\ga_5$ is defined in 4 dimensional space, i.e.,
$\ga_5=-i\ga^0\ga^1\ga^2\ga^3$. Then, following identity holds
\begin{align}
\ga_5\ga^-\s{s}_{a\perp}=i\ga^-\tilde{s}_{a\perp}\cdot\ga,\
\tilde{s}_{a\perp}^\mu=\ep^{-+\mu\rho}s_{a\perp\rho}.
\label{eq:ga5_id}
\end{align}
This means $\ga_5$ in spin projection operators can be eliminated. Then, partonic
cross section can be written as
\begin{align}
d\Delta\hat{\sig}=\tilde{s}_{a\perp}^i \tilde{s}_{b\perp}^j W_{ij}(k_a,k_b,p_1),
\end{align}
where $i,j$ denote Lorentz indices in 4 dimensional space and $\mu,\nu$ denote Lorentz indices in n dimensional space. Although $W_{ij}$ is defined in 4 dimensional space,
we can still calculate it in n dimensional space first and then project the result
to 4 dimensional space, i.e., $W_{ij}=g_{i\mu}g_{j\nu}W^{\mu\nu}$. On the other hand, in the scheme with anti-commuting $\ga_5$, i.e. $\{\ga_5,\ga^\mu\}=0$,
the same tensor $W_{\mu\nu}$ can be obtained after one $\ga_5$ in spin projection operators is exchanged with other gamma matrices and then is eliminated by another
$\ga_5$ due to $\ga_5\ga_5=1$. This is our proof for
the equivalence of the two $\ga_5$ schemes
for calculations involving transversity PDF. The proof may also help
to understand the results of \cite{Mukherjee:2003pf,Mukherjee:2005rw},
where pion and prompt photon as probes of
transversity on hadron colliders are explicitly calculated in the two mentioned $\ga_5$ schemes and the same cross sections are obtained. Actually, to
all orders of $\al_s$, the two schemes produce the same result.
Of course, for scatterings involving other polarized PDFs, like helicity PDF,
for which the spin projection operator is $\ga_5\ga^\pm$, we cannot eliminate $\ga_5$ as done in eq.(\ref{eq:ga5_id}),
and we expect the corresponding results in HVBM scheme and anti-commuting scheme are different. Generally, even though there are two $\ga_5$ in a Dirac trace, we still
have no reason to claim that HVBM and anti-commuting schemes will generate the same
result. For example, $\ga^\mu\ga_5\ga_\mu\ga_5=-4-\ep,-4+\ep$ in HVBM scheme and in
anti-commuting scheme, respectively.

\section{Numerical results}
In this section, we will present our numerical results for $A_{TT}$ defined in eq.(\ref{eq:ATT_def}), with azimuthal angle integrated over. We consider the heavy quark production on both proton-proton collider and proton-antiproton collider.
We know that beyond
tree level, the production cross section for heavy quark is different from
that for heavy anti-quark. Thus, in the following we will also
discuss charge average $d\sig_{ave}$ and charge asymmetry $d\sig_{asy}$ for
both unpolarized and polarized cross sections, which are defined as
\begin{align}
d\sig_{ave}\equiv \frac{d\sig^Q+d\sig^{\bar{Q}}}{2},\
d\sig_{asy}\equiv \frac{d\sig^Q-d\sig^{\bar{Q}}}{2}.
\label{eq:charge_ave}
\end{align}
Correspondingly, we define two azimuthal asymmetries $A_{TT}^{ave}$ and
$A_{TT}^{asy}$ according to these two different cross sections, respectively. By default,
$A_{TT}$ in the following is $A_{TT}^{ave}$. In the following, leading order(LO)
result is tree level result, and next-to-leading order(NLO) result includes tree
and one-loop results. Polarizations of (anti-)proton beams are assumed to be one,
i.e., $|s_{a\perp}|=|s_{b\perp}|=1$.

\begin{figure}
\begin{center}
\begin{minipage}[b]{0.23\textwidth}
\includegraphics[width=\textwidth]{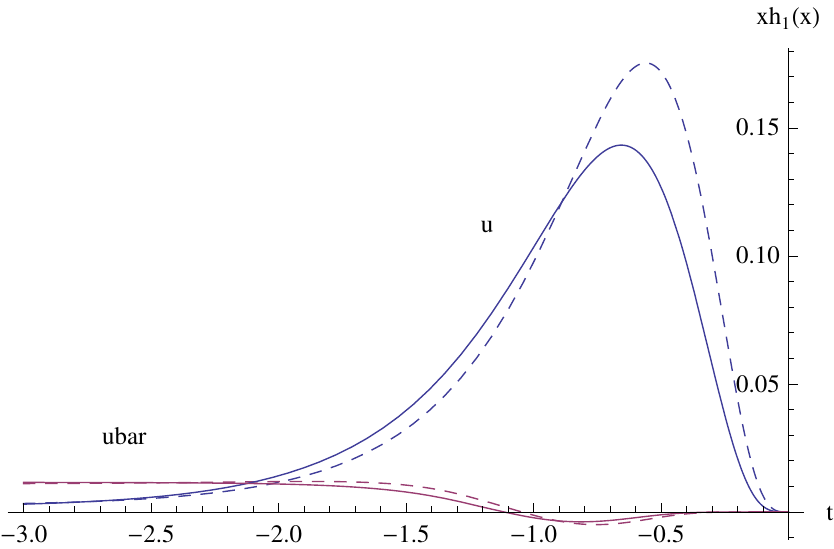}
(a)
\end{minipage}
\begin{minipage}[b]{0.23\textwidth}
\includegraphics[width=\textwidth]{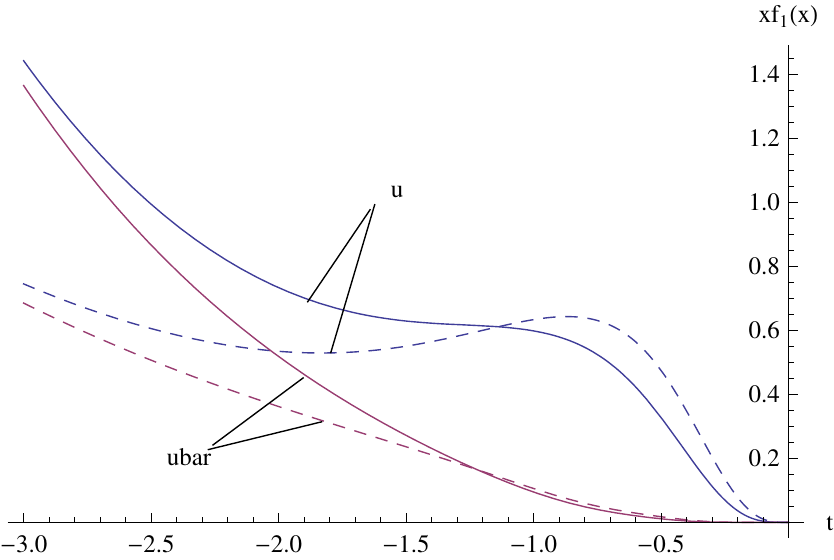}
(b)
\end{minipage}
\begin{minipage}[b]{0.23\textwidth}
\includegraphics[width=\textwidth]{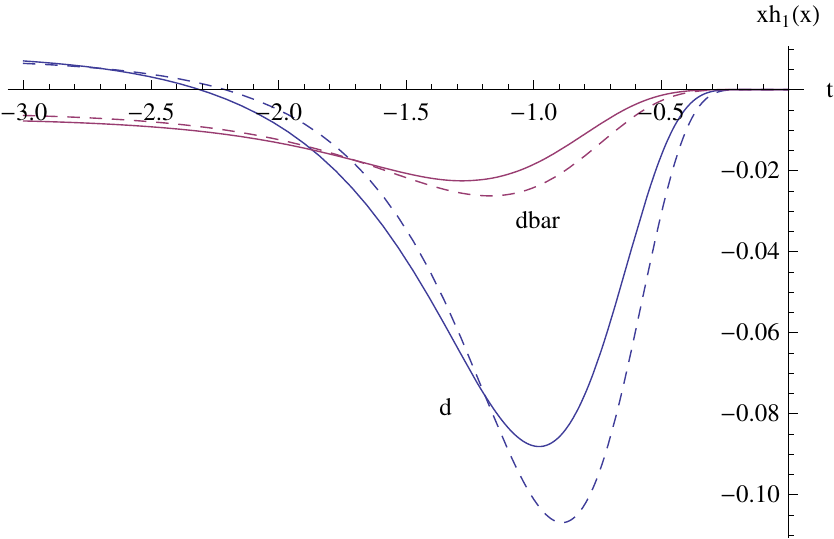}
(c)
\end{minipage}
\begin{minipage}[b]{0.23\textwidth}
\includegraphics[width=\textwidth]{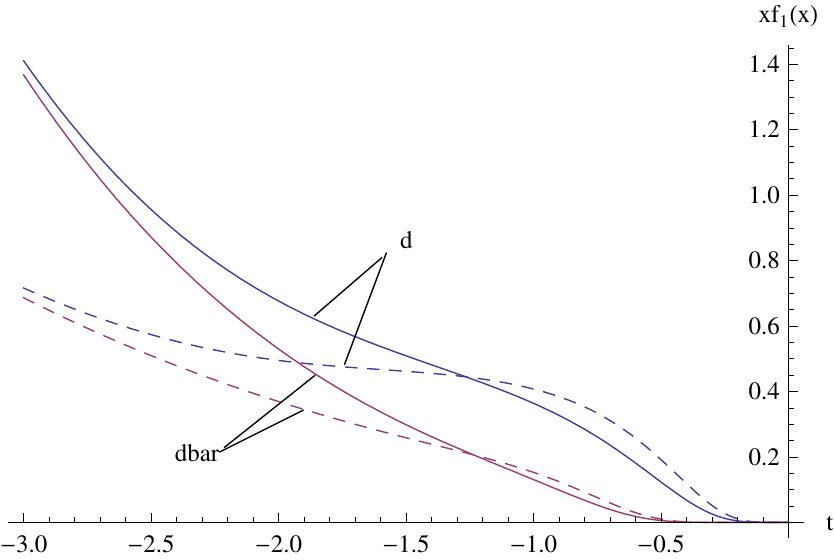}
(d)
\end{minipage}
\end{center}
\caption{Tansversity PDFs and unpolarized PDFs for u($\bar{u}$) and d($\bar{d}$) quarks at two different
scales in our model. In each diagram, the dashed and real lines stand for PDFs at
$\mu=5\gev$ and $60\gev$, respectively. (a) and (c) are for
transversity PDFs, while (b) and (d) are for unpolarized
PDFs. $x=10^t$ in each diagram.}
\label{fig:PDF}
\end{figure}

\begin{figure}
\begin{center}
\begin{minipage}[b]{0.3\textwidth}
\includegraphics[width=\textwidth]{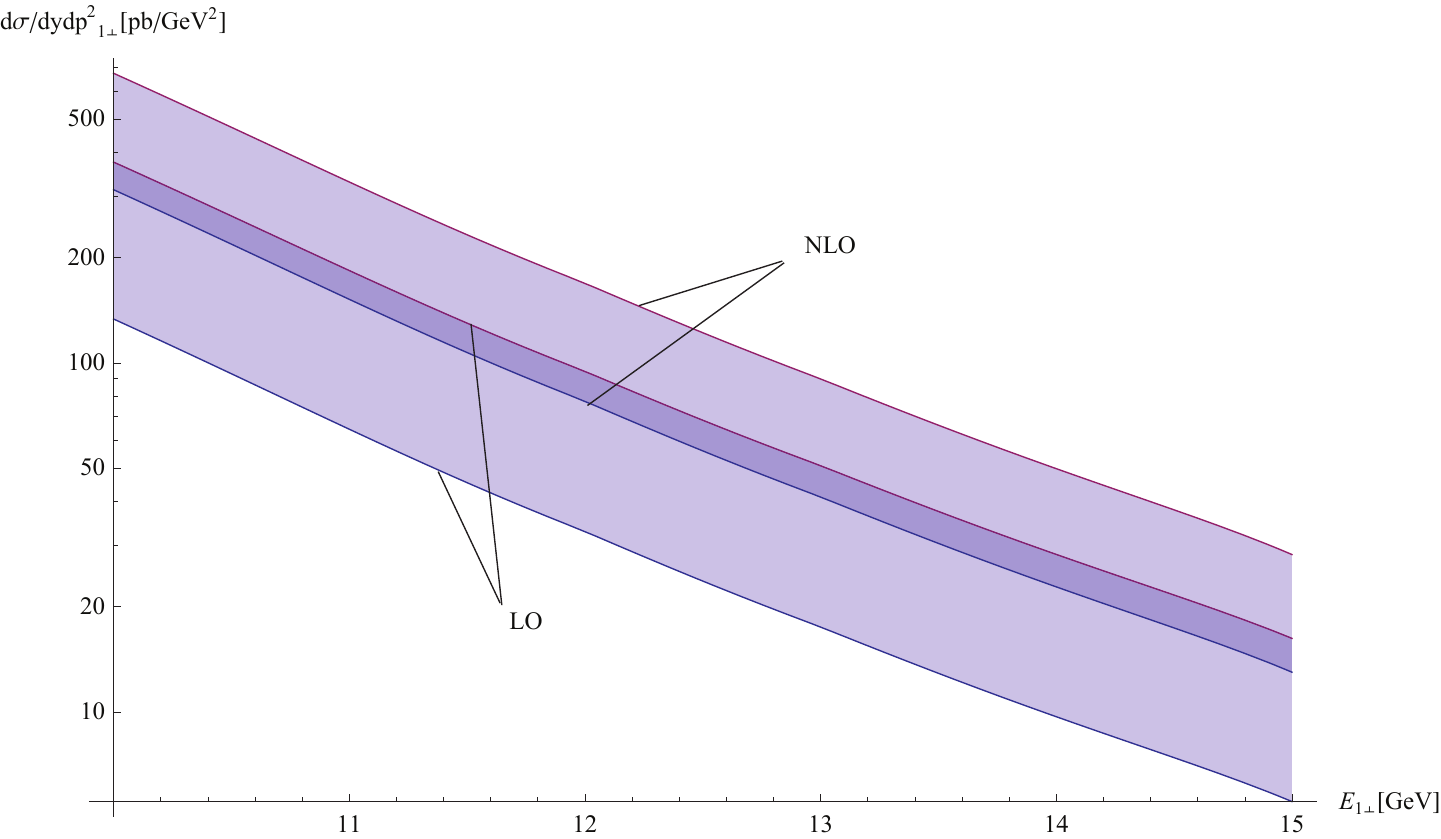}
(a)
\end{minipage}
\begin{minipage}[b]{0.3\textwidth}
\includegraphics[width=\textwidth]{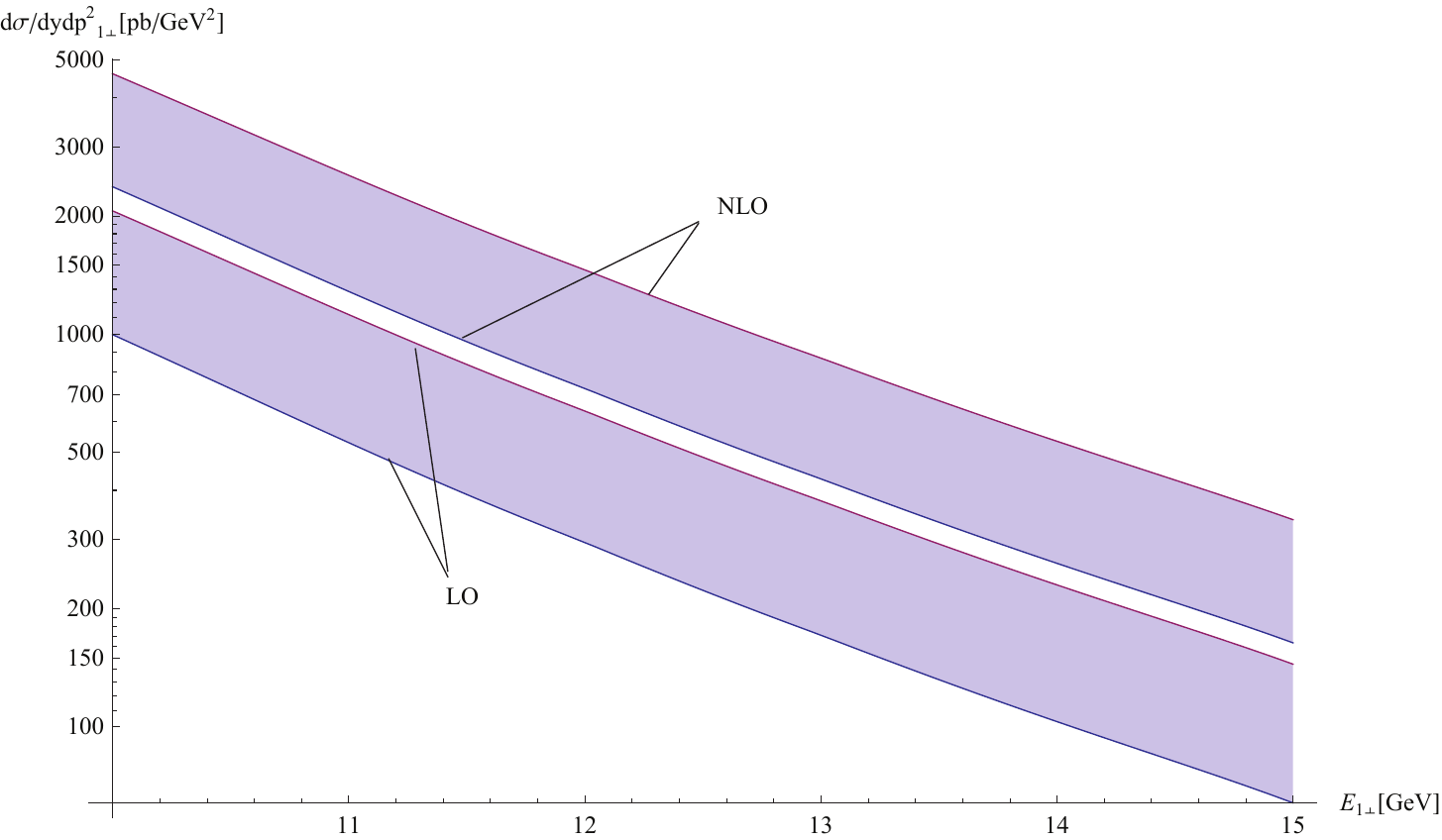}
(b)
\end{minipage}
\begin{minipage}[b]{0.3\textwidth}
\includegraphics[width=\textwidth]{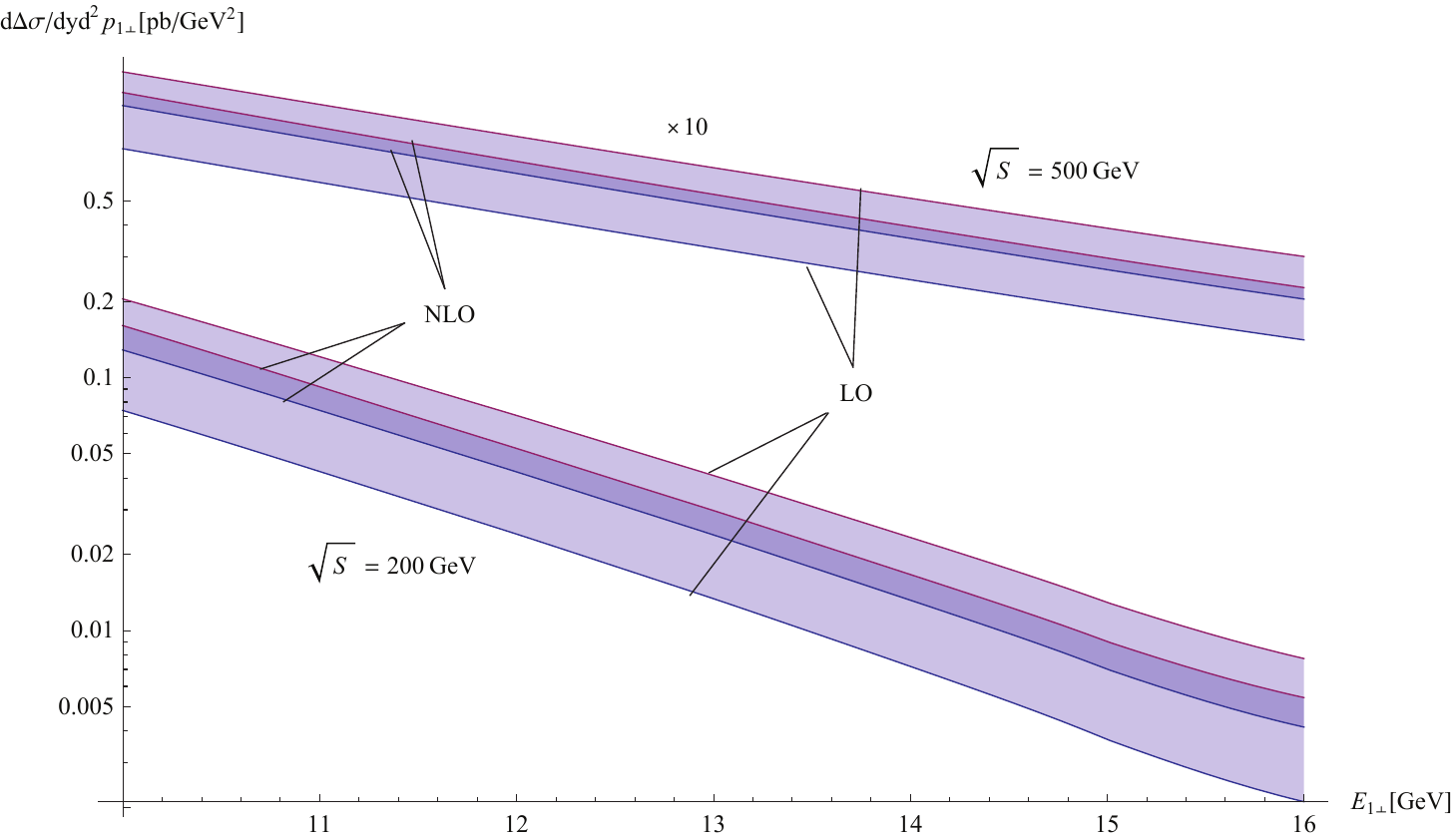}
(c)
\end{minipage}
\end{center}
\caption{Unpolarized and polarized cross sections for bottom production on RHIC.
(a) and (b) are the unpolarized cross sections for $\sqrt{S}=200$
and $500\gev$, respectively. (c) is the polarized cross section.
In all of these cross sections, we have set rapidity and azimuthal angle of
heavy quark to be zero, i.e., $y=0,\phi=0$. The uncertainty band is obtained by
varying renormalization scale $\mu$ from $E_{1\perp}/2$ to $2E_{1\perp}$.
Note that in (a), LO and NLO bands have an overlap.}
\label{fig:cross_section}
\end{figure}

To calculate the cross sections, we need unpolarized PDFs and transversity PDFs.
In this work, the unpolarized PDFs are taken as MSTW2008 PDFs\cite{Martin:2009iq}. For transversity PDFs, the extracted valence quark transversity PDFs from either TMD formalism
or Di-hadron formalism are in agreement with each other within uncertainty range.
Thus, we take the extracted transversity PDFs for $u,d$ quarks in \cite{Kang:2015msa} as reference. The remaining sea quark transversity PDFs for $\bar{u},\bar{d}, \bar{s}$ are still absent in literature, and are assumed to be the same as corresponding sea quark helicity PDFs
at a certain low energy scale\cite{Barone:2005cr,deFlorian:2017ogw,Soffer:2002tf}.
In our case, the low energy scale is
$\mu^2=2.4\gev^2$, which is the starting scale of \cite{Kang:2015msa} for the extraction of transversity PDFs. The helicity PDFs are taken as DSSV type\cite{deFlorian:2009vb,deFlorian:2014yva}.
Because unpolarized PDFs are always greater than helicity PDFs for most momentum
fraction x, Soffer's bound\cite{Soffer:1994ww} is always satisfied.
Moreover, for unpolarized cross section we use NLO PDFs and NLO $\al_s$, but for
polarized cross section we just use LO transversity PDFs and LO $\al_s$.
Compared with the evolution of unpolarized PDFs, the scale dependence of transversity PDFs is very small, as shown in Fig.\ref{fig:PDF}.
From Fig.\ref{fig:cross_section},
the scale uncertainty
of polarized cross section by varying renormalization scale $\mu$ from $E_{1\perp}/2$ to $2E_{1\perp}$ is much smaller than the scale uncertainty of
unpolarized cross section. Thus, we think it is sufficient to use LO transversity
PDFs to estimate polarized cross sections.

\begin{figure}
\begin{center}
\begin{minipage}{0.45\textwidth}
\includegraphics[width=\textwidth]{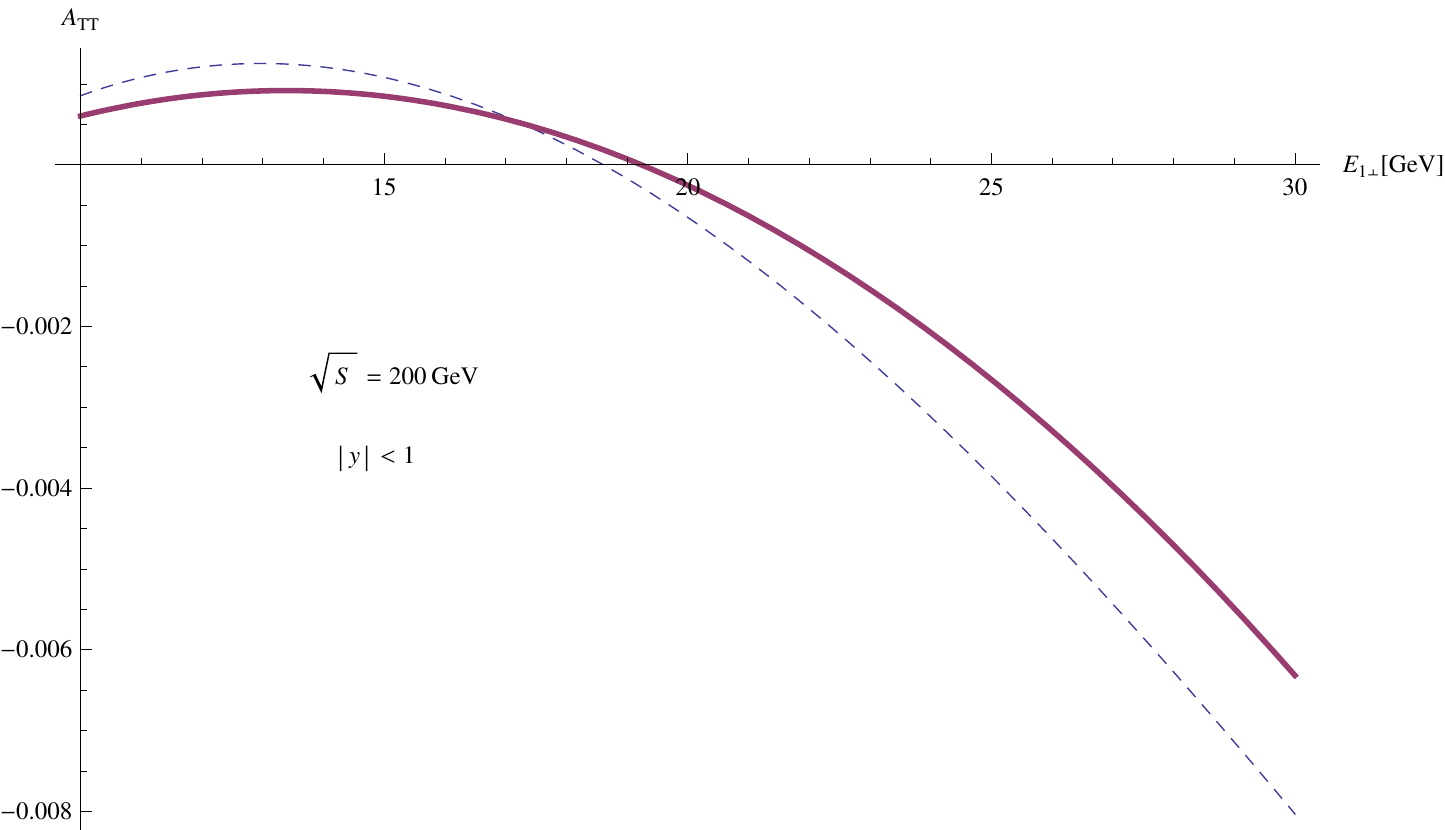}
(a)
\end{minipage}
\begin{minipage}{0.45\textwidth}
\includegraphics[width=\textwidth]{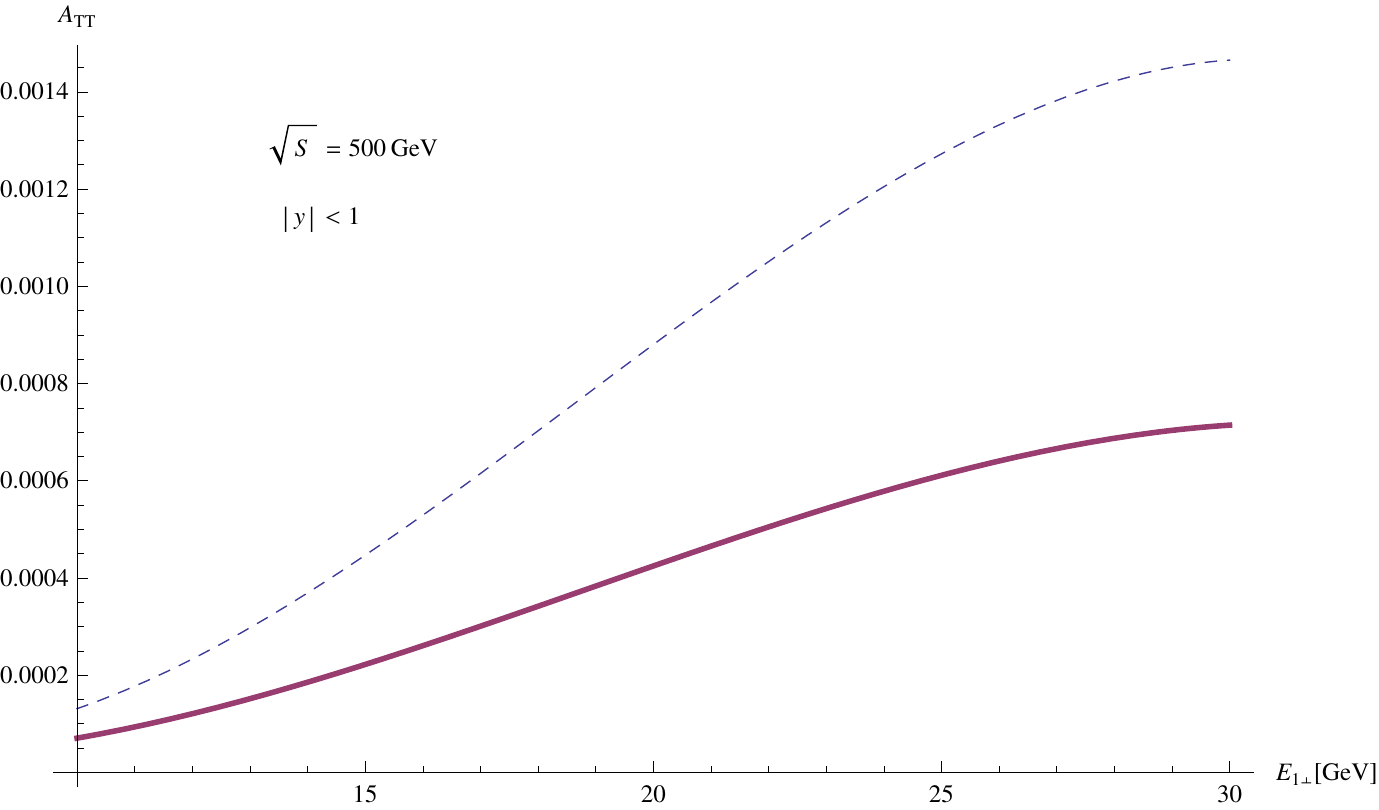}
(b)
\end{minipage}
\end{center}
\caption{Azimuthal asymmetries of heavy quark(bottom), in which heavy quark rapidity is integrated over $(-1,1)$. The dashed line is LO result and the real line is NLO result. The curves are obtained by setting $\mu=E_{1\perp}$. }
\label{fig:ATT}
\end{figure}
For bottom production on proton-proton colliders, such as RHIC\cite{RHIC-spin:2013woa} with $\sqrt{S}=200
$, $500\gev$, the cross sections are given in Fig.\ref{fig:cross_section}.
The hard coefficients
of unpolarized cross section are taken from \cite{Nason:1989zy}.
From LO to NLO, the
corrections to unpolarized cross sections are large, which can be greater than
$100\%$, and the scale uncertainty by varying renormalization scale $\mu$
is not reduced significantly. On the
contrary, the scale uncertainty of polarized cross sections is reduced greatly
from LO to NLO, as shown in Fig.\ref{fig:cross_section}(c).
Note that the central values of LO and
NLO polarized cross sections are very close to each other in this figure,
which means
the loop correction to polarized cross section is not large. Thus, we conclude
the main correction to $A_{TT}$ and the main scale uncertainty are caused by
unpolarized cross section. For
more precise estimate, one has to use NNLO unpolarized cross section in
the denominator of $A_{TT}$.

As expected, on RHIC due to the smallness of sea transversities, the azimuthal
asymmetry $A_{TT}$ is very small. When $E_{1\perp}\leq 15\gev$, $A_{TT}$ is
of order $10^{-4}$. One can see this
from Fig.\ref{fig:ATT}.
According to the estimate of \cite{Soffer:2002tf},
the observable asymmetry on RHIC should at least be larger than $10^{-3}$.
Thus, the observation of $A_{TT}$ on RHIC is very difficult.

\begin{figure}
\begin{center}
\includegraphics[scale=0.5]{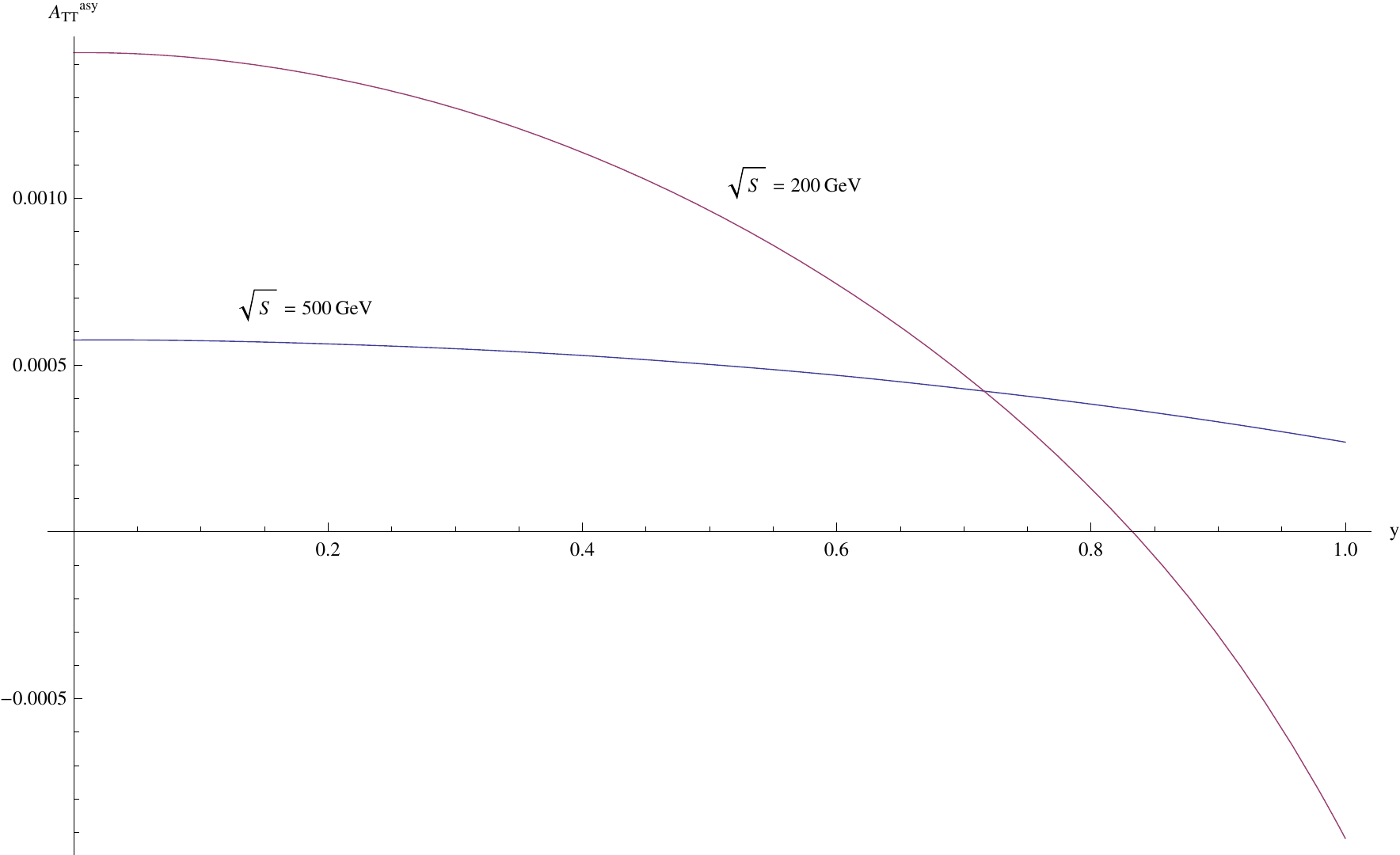}
\end{center}
\caption{Azimuthal asymmetry related to heavy quark charge asymmetry, i.e., $A_{TT}^{asy}$. $E_{1\perp}=10\gev$ and $\mu=E_{1\perp}$ are taken
in the calculation.  }
\label{fig:ATT_AS}
\end{figure}
Besides the smallness of sea transversities, the very large contribution of gluon
to unpolarized cross section also suppresses $A_{TT}$(or $A_{TT}^{ave}$
explicitly). However, for heavy quark production, the charge asymmetry does
not receive contribution from gluon-gluon scattering due to charge conjugation
and Bose symmetries. Thus, one may guess $A_{TT}^{asy}$ will be sizable. But
this is not the case. For $E_{1\perp}=10\gev$, $A_{TT}^{asy}$ is shown in
Fig.\ref{fig:ATT_AS} for both cases with $\sqrt{S}=200$, $500\gev$ and
$E_{1perp}=10\gev$. Only for
$\sqrt{S}=200\gev$, $A_{TT}^{asy}$ can reach $10^{-3}$ in very central region(
$|y|\leq0.2$).
We note that
for unpolarized cross section, $d\sig^{asy}$ is nearly 2 orders smaller than
$d\sig^{ave}$ in the considering kinematical region. Interestingly, for
poarized cross section, $d\Delta\sig^{asy}$ is also smaller than
$d\Delta\sig^{ave}$ by about 2 orders.
Thus, $A_{TT}^{asy}$
actually is of the same order as $A_{TT}^{ave}$, and can hardly give us more
information about transversity.

Since polarized anti-proton beam is not available now, there is no experiment for DSA on proton-antiprtonton colliders.
But an interesting polarized anti-proton program was proposed by PAX collaboration of FAIR at GSI\cite{GSI:2004dza}. The main purpose is to measure the lepton angular
distribution in double transversely polarized Drell-Yan.
There are collider and fixed target schemes. In collider scheme, the momentum of
polarized anti-proton can reach $15\gev/c$, and the momentum of polarized proton can reach
$3.5\gev/c$. In fixed target scheme, the momentum of polarized anti-proton can
reach $22\gev/c$. In collider scheme, charm can be produced. Thus,
it is interesting to see whether $A_{TT}$ for charm production can
be measured on GSI.
Here we consider the charge average only. The resulting $A_{TT}$ is shown in
the tables in Appendix.\ref{sec:GSI} in various kinematical regions.
From these results, $A_{TT}$ on GSI
is above $1\%$, and in some regions it can be greater than $10\%$.
Thus, measuring $A_{TT}$ for charm production on future GSI experiments will be
very helpful to determine or justify the extracted valence transversity PDFs.

\section{Summary}
In this work, we calculate one-loop QCD correction to single heavy quark inclusive production on double transversely polarized hadron colliders. Analytic results are given. The tensor integrals appearing in both virtual and real corrections are
treated similarly and are reduced by FIRE into several master integrals.
For real correction, the soft and collinear divergences of master integrals
can be separated easily. Then, these master integrals are calculated with
Feynman parameters. The results are the same as those in literature\cite{Beenakker:1988bq,Beenakker:1990maa}.
As a check, we also use our program to calculate the unpolarized cross section with $q\bar{q}$ as partons entering hard scattering, and numerically, the obtained hard coefficients are
the same as known results in literature\cite{Nason:1989zy,Beenakker:1990maa}. With the analytic results, numerical
estimates on proton-proton collider(RHIC,$\sqrt{S}=200,500\gev$) and proton-antiproton collider(e.g.,GSI, collider scheme) are given. The azimuthal
asymmetry $A_{TT}$ for bottom production on RHIC are suppressed by the smallness of sea transversity PDF and by dominated gluon contribution to unpolarized cross section. The resulting $A_{TT}$ on RHIC is of order $10^{-4}$, which is too small to be measured. Even with charge asymmetry of heavy quark production taken into account, the related asymmetry $A_{TT}^{asy}$ is not enhanced greatly.
On GSI, valence transversity PDFs give main contribution to $A_{TT}$. For charm
production, $A_{TT}$ can be greater than $10\%$, which provides a good chance
to extract or justify the extracted valence transversity PDFs. Moreover,
the main uncertainty of $A_{TT}$ is caused by the large scale dependence of
unpolarized cross sections. To get more precise estimates, NNLO unpolarized
cross sections should be used.

\section*{Acknowledgements}
The hospitality of USTC during the completion of this paper is appreciated. This work is supported by National Nature Science Foundation of China(NSFC) with contract No.11605195.

\appendix
\section{Bubble and tadpole integrals}\label{sec:virtual_integrals}
The bubble and tadpole integrals are defined as
\begin{align}
\mu^{4-n}\int \frac{d^n l}{(2\pi)^n}\frac{1}{[l^2-m_1^2][(l+p)^2-m_2^2]}
=&\frac{i}{16\pi^2}\frac{(4\pi)^{\ep/2}}{\Gamma(1-\ep/2)}b(p^2,m_1,m_2),\no
\mu^{4-n}\int \frac{d^n l}{(2\pi)^n}\frac{1}{l^2-m^2}=&
\frac{i}{16\pi^2}\frac{(4\pi)^{\ep/2}}{\Gamma(1-\ep/2)} a_0(m).
\end{align}
The integrals are expanded to $O(\ep)$. The calculation is done in
physical region, with $s>0,t<0$.

\begin{enumerate}
\item{}
\begin{align}
b(t,m,0)=&\Big(\frac{\mu^2}{m^2}\Big)^{\ep/2}
 \Big(\frac{m^2-t}{m^2}\Big)^{-\ep/2}
 B(1-\ep/2,\ep/2)J\no
 J=& 1-\frac{\ep}{2}\Big[
 \frac{m^2 \log
   \left(\frac{m^2}{m^2-t}\right)}{t}-2
 \Big]\no
 &+\frac{\ep^2}{8}\frac{1}{3t}\Big[
 -6 \left(m^2-t\right)
   \text{Li}_2\left(\frac{m^2}{m^2-t}\right)+3
   m^2 \log ^2\left(\frac{m^2}{m^2-t}\right)\no
   &-6\log \left(\frac{m^2}{m^2-t}\right)
   \left(\left(m^2-t\right) \log
   \left(\frac{t}{t-m^2}\right)+2
   m^2\right)+\pi ^2 m^2-\pi ^2 t+24 t
 \Big],
\end{align}
\item{}
\begin{align}
b(s,0,0)=&\Big(\frac{\mu^2}{m^2}\Big)^{\ep/2}
B(1-\frac{\ep}{2},\frac{\ep}{2})B(1-\frac{\ep}{2},1-\frac{\ep}{2})
\Big[
1-\frac{\ep}{2}\ln\frac{s}{m^2}+\frac{\ep^2}{8}(\ln^2\frac{s}{m^2}-\pi^2)
\Big],
\end{align}
\item{}
\begin{align}
b(s,m,m)=& \lrb{\frac{\mu^2}{m^2}}^{\ep/2}B(1-\frac{\ep}{2},\frac{\ep}{2})\no
&\times
\Big[
1-\frac{\ep}{2}\int_0^1 dx \ln|1+A x(1-x)|+\frac{\ep^2}{8}\Big(
-\pi^2\sqrt{1+4/A}+
\int_0^1 dx
\ln^2|1+Ax(1-x)|
\Big)
\Big],\no
A=&\frac{-s}{m^2},
\end{align}
\item{}
\begin{align}
a_0(m)=& -\Gamma(1-\frac{\ep}{2})\Gamma(-1+\frac{\ep}{2})(m^2)^{1-\frac{\ep}{2}}
=\Big(\frac{\mu^2}{m^2}\Big)^{\ep/2}\frac{m^2}{1-\ep/2}
B(1-\ep/2,\ep/2)\no
=&\Big(\frac{\mu^2}{m^2}\Big)^{\ep/2}m^2\Big[
\frac{2}{\epsilon }+1
+\frac{1}{12} \left(6+\pi ^2\right) \epsilon
\Big].
\end{align}
\end{enumerate}

\section{Real integrals}\label{sec:Ri}
The real integrals defined in eq.(\ref{eq:Ri}) are given here.  $R_3$ is calculated to $O(1)$, and $R_6$ is calculated to $O(\ep^2)$. Others are calculated to $O(\ep)$.
All are compared with the formulas in \cite{Beenakker:1988bq}. Numerically,
the two results are precisely the same.
$R_i$ are organized as follows:
\begin{align}
R_i=&N_\ep\Big[\frac{2}{\ep} R_i^{(-1)}+R_i^{(0)}+\frac{\ep}{2} R_i^{(1)}\Big],\
N_\ep=\frac{2\pi^{1-\frac{\ep}{2}}}{\Gamma(1-\frac{\ep}{2})}.
\end{align}
The explicit expressions are
\begin{enumerate}
\item{$R_1(w):$}
\begin{align}
R_1^{(-1)}=&-4 w^2;\no
R_1^{(0)}=& -8 w^2 \log (w);\no
R_1^{(1)}=&2 w^2 \Big[4 \text{Li}_2(1-2 w)-2
   \text{Li}_2(-2 w)-2
   \text{Li}_2\left(\frac{1}{2
   w+1}\right)\no
&-\log ^2(2 w+1)+2 \log (w) \log
   \left(\frac{32}{2 w+1}\right)-2 \log (2)
   \log (w (2 w+1))+\frac{\pi ^2}{3}+4 \log
   ^2(2)\Big];
\end{align}

\item{$R_2(\de,w):$}
\begin{align}
R_2^{(-1)}=&-\frac{2 w^2}{2 w^2 (\delta -1)+1};\no
R_2^{(0)}=&\frac{2 w^2 \log \left(\frac{\left(2 w^2
   (\delta -1)+1\right)^2}{w^4 \left(\delta
   ^2-1\right)}\right)}{2 w^2 (\delta -1)+1};\no
R_2^{(1)}=&-\frac{w^2}{3 \left(2 w^2 (\delta
   -1)+1\right)}
    \Big[6 \text{Ir2}(\de,w)-6 \text{Li}_2\left(\frac{w
   (\delta -1)+1}{2 (\delta -1)
   w^2+1}\right)\no
&+3 \log \left(\frac{w (2
   w+1)}{2 w^2 (\delta -1)+1}\right) \left(3
   \log \left(\frac{w}{2 w^2 (\delta
   -1)+1}\right)+2 \log ((2 w-1) (\delta
   -1))+\log \left(\frac{1}{4} (2
   w+1)\right)\right)\no
&-6 \log (2) \log
   \left(\frac{w (2 w-1) (\delta -1)}{2 w^2
   (\delta -1)+1}\right)\no
&-3 \log \left(\frac{2
   w^2 (\delta +1)}{4 w^2-1}\right) \left(\log
   \left(16 (w (-\delta )+w+1)^2\right)-2 \log
   (w (\delta -1)+1)\right)+\pi ^2+3 \log
   ^2(2)\Big];\no
\text{Ir2}(\de,w)=&
\int_0^1 dz \frac{\xi_3-\xi_2}{(z+\xi_3)(z+\xi_2)}
\ln\frac{(1-z)(z+\xi_1)}{(z+\xi_2)|z+\xi_3|}
-\int_0^1 dz \frac{1+\xi_1}{z(1+\xi_1-z)}
\ln\frac{(1-\frac{z}{1+\xi_2})(1-\frac{z}{1+\xi_3})}{1-\frac{z}{1+\xi_1}};\no
\xi_1=&\frac{w(2w-1)(\Delta-1)}{1+w(\Delta-1)},\
\xi_2=2w-1,\
\xi_3=\frac{w(1+\Delta)}{1+w-w\Delta}.
\label{eq:Ir2}
\end{align}

\item{$R_3(\de,\Delta,w):$}
\begin{align}
R_3^{(-1)}=&0;\no
R_3^{(0)}=& \frac{2 w^2}{\sqrt{4 w^2
   \left(w^2 (\delta -\Delta )^2+\delta
   \Delta -1\right)+1}} \log \left(-\frac{2 w^2 (\delta
   \Delta -1)+\sqrt{4 w^2 \left(w^2 (\delta
   -\Delta )^2+\delta  \Delta
   -1\right)+1}+1}{w^2 (2-2 \delta  \Delta
   )+\sqrt{4 w^2 \left(w^2 (\delta -\Delta
   )^2+\delta  \Delta
   -1\right)+1}-1}\right);\no
\end{align}

\item{$R_4:$}
\begin{align}
R_4^{(-1)}=&-1;\no
R_4^{(0)}=& 2 \log (2);\no
R_4^{(1)}=&2 \left(\frac{\pi ^2}{12}-\log ^2(2)\right);
\end{align}

\item{$R_5(\de):$}
\begin{align}
R_5^{(-1)}=&0;\no
R_5^{(0)}=&\log \left(\frac{\delta +1}{\delta -1}\right) ;\no
R_5^{(1)}=&-\text{Li}_2\left(-\frac{2}{\delta
   -1}\right)+\text{Li}_2\left(\frac{2}{\delta
   +1}\right)-2 \log (2) \log
   \left(\frac{\delta +1}{\delta -1}\right);
\end{align}

\item{$R_6:$}
\begin{align}
R_6=&\int d\Omega_{n-1}=N_\ep \int_0^\pi d\theta \sin^{n-3}{\theta}
=N_\ep 2^{1-\ep} B(1-\frac{\ep}{2},1-\frac{\ep}{2})\no
=&
N_\ep \Big[2+\epsilon  (2-\log (4))+\epsilon ^2 \left(2-\frac{\pi ^2}{12}
+\log^2(2)-\log (4)\right)+O\left(\epsilon ^3\right)\Big].
\end{align}
\end{enumerate}

\newpage
\section{Numerical results on GSI}\label{sec:tables}
Numerical results for charm production on GSI are listed in following tables.
In the unpolarized cross sections
the azimuthal angle of heavy quark is integrated over, and we show the results for
\begin{align}
\frac{d\sig^{unp}}{dy d\vec{p}_{1\perp}^2}=\pi\frac{d\sig^{unp}}{dy d^2 p_{1\perp}},
\end{align}
where the differential cross section on right-hand side is defined in eq.(\ref{eq:hadron_cross}).
Further, the contributions from different parton flavors in initial state of subprocess are also shown in the following tables. That is,
\begin{align}
\frac{d\sig^{unp}}{dy d\vec{p}_{1\perp}^2}=&
\sum_{(ij)=gg,qg,q\bar{q}}
\frac{d\sig_{ij}^{unp}}{dy d\vec{p}_{1\perp}^2}.
\end{align}
$d\sig_{gg}$ is the cross section given by subprocess $gg\rightarrow Q+X$;
$d\sig_{qg}$ is the cross section given by subprocesses $qg\rightarrow Q+X$ and
$gq\rightarrow Q+X$, with $q=u,d,s,\bar{u},\bar{d},\bar{s}$;
$d\sig_{q\bar{q}}$ is the cross section given by subprocesses $q\bar{q}\rightarrow
Q+X$, with $q\bar{q}=u\bar{u},d\bar{d},s\bar{s},\bar{u}u,\bar{d}d,\bar{s}s$.
For convenience, following notations are introduced:
\begin{align}
\Sigma_{ij}\equiv \frac{d\sig_{ij}^{unp}}{dy d\vec{p}_{1\perp}^2},\
\Sigma \equiv \sum_{ij}\Sigma_{ij}=\frac{d\sig^{unp}}{dy d\vec{p}_{1\perp}^2},\
\Sigma_T \equiv \frac{d\Delta \sig}{dy d^2 p_{1\perp}}
\Big|_{\phi=0,|s_{a\perp}|=|s_{b\perp}|=1}.
\end{align}
For polarized cross section, it is clear that $d\Delta\sig=d\Delta\sig_{q\bar{q}}$,
and quark flavors $q\bar{q}$ are the same as those of unpolarized cross section.
In polarized cross section we have set the azimuthal angle of heavy quark to be
zero, and assumed the polarizations of initial (anti-)proton beams to be 1. With
these notations, the azimuthal asymmetry $A_{TT}=\frac{\pi}{2}\frac{\Sigma_T}{\Sigma}$.

Moreover, as illustrated in text, all cross sections here are for the charge average
of heavy quark, i.e. $d\sig_{ave}$ defined in eq.(\ref{eq:charge_ave}). The renormalization scale dependence($\mu$ dependence) is also calculated. As done in \cite{Nason:1989zy}, each cross section
is calculated by setting $\mu=\mu_0/2,\mu_0,2\mu_0$ with $\mu_0=E_{1\perp}$.
In following tables, corresponding to every $(E_{1\perp},y)$,
the central number is given by $\mu=\mu_0$, while the upper
and lower numbers are given by $\mu=\mu_0/2,2\mu_0$, respectively.
For the case $E_{1\perp}=3\gev$, $\mu_0/2$ is replaced
by $\sqrt{2.4}\gev$. The unit of $E_{1\perp}$ is GeV and the unit of
cross section is pb$/\gev^2$. Charm mass $m_c=1.40\gev$ and bottom mass $m_b=4.75\gev$. All calculations are performed in $\msb$ scheme.

\subsection{Results on GSI $S=216.4\gev^2$}\label{sec:GSI}
\begin{tabular}[c]{|c|p{2.4cm}|p{2.4cm}|p{2.4cm}|p{2.4cm}|p{2.4cm}|p{2.0cm}|}
\hline
GSI&&&&&&\\
$S=216.4\gev^2$&  $\Sigma_{gg}(pb/\gev^2)$     &  $\Sigma_{qg}(pb/\gev^2)$     &  $\Sigma_{q\bar{q}}(pb/\gev^2)$   &  $\Sigma(pb/\gev^2)$  & $\Sigma_T(pb/\gev^2)$&
$A_{TT}$\\
$E_{1\perp}=3\gev$&&&&&&\\
\hline
& 2.59$\times 10^3$ & 2.60$\times 10^2$ & 8.52$\times 10^3$ & 1.14$\times 10^4$ & 1.58$\times 10^2$ & 2.18$\times 10^{-2}$\\
$y=0.0$ & 8.64$\times 10^2$ & -2.22$\times 10$ & 5.73$\times 10^3$ & 6.58$\times 10^3$ & 1.79$\times 10^2$ & 4.26$\times 10^{-2}$\\
& 3.02$\times 10^2$ & -2.26$\times 10$ & 3.22$\times 10^3$ & 3.50$\times 10^3$ & 1.17$\times 10^2$ & 5.24$\times 10^{-2}$\\
\hline
& 2.36$\times 10^3$ & 2.09$\times 10^2$ & 7.15$\times 10^3$ & 9.72$\times 10^3$ & 1.33$\times 10^2$ & 2.16$\times 10^{-2}$\\
$y=0.3$ & 7.53$\times 10^2$ & -2.09$\times 10$ & 4.80$\times 10^3$ & 5.53$\times 10^3$ & 1.48$\times 10^2$ & 4.20$\times 10^{-2}$\\
& 2.57$\times 10^2$ & -1.93$\times 10$ & 2.67$\times 10^3$ & 2.91$\times 10^3$ & 9.55$\times 10$ & 5.15$\times 10^{-2}$\\
\hline
& 1.66$\times 10^3$ & 1.01$\times 10^2$ & 4.02$\times 10^3$ & 5.78$\times 10^3$ & 7.66$\times 10$ & 2.08$\times 10^{-2}$\\
$y=0.6$ & 4.67$\times 10^2$ & -1.50$\times 10$ & 2.66$\times 10^3$ & 3.11$\times 10^3$ & 7.90$\times 10$ & 3.99$\times 10^{-2}$\\
& 1.47$\times 10^2$ & -1.09$\times 10$ & 1.44$\times 10^3$ & 1.58$\times 10^3$ & 4.92$\times 10$ & 4.89$\times 10^{-2}$\\
\hline
& 7.18$\times 10^2$ & 2.03$\times 10$ & 1.22$\times 10^3$ & 1.96$\times 10^3$ & 2.40$\times 10$ & 1.93$\times 10^{-2}$\\
$y=0.9$ & 1.59$\times 10^2$ & -5.66  & 7.76$\times 10^2$ & 9.30$\times 10^2$ & 2.15$\times 10$ & 3.64$\times 10^{-2}$\\
& 4.27$\times 10$ & -3.02  & 4.01$\times 10^2$ & 4.41$\times 10^2$ & 1.24$\times 10$ & 4.43$\times 10^{-2}$\\
\hline
& 9.02$\times 10$ & 3.49$\times 10^{-2}$ & 9.70$\times 10$ & 1.87$\times 10^2$ & 1.91  & 1.60$\times 10^{-2}$\\
$y=1.2$ & 1.27$\times 10$ & -4.39$\times 10^{-1}$  & 5.46$\times 10$ & 6.69$\times 10$ & 1.29  & 3.03$\times 10^{-2}$\\
& 2.49  & -1.51$\times 10^{-1}$  & 2.51$\times 10$ & 2.74$\times 10$ & 6.38$\times 10^{-1}$  & 3.66$\times 10^{-2}$\\
\hline
\end{tabular}

\begin{tabular}[c]{|c|p{2.4cm}|p{2.4cm}|p{2.4cm}|p{2.4cm}|p{2.4cm}|p{2.0cm}|}
\hline
GSI&&&&&&\\
$S=216.4\gev^2$&  $\Sigma_{gg}(pb/\gev^2)$     &  $\Sigma_{qg}(pb/\gev^2)$     &  $\Sigma_{q\bar{q}}(pb/\gev^2)$   &  $\Sigma(pb/\gev^2)$  & $\Sigma_T(pb/\gev^2)$&
$A_{TT}$\\
$E_{1\perp}=4\gev$&&&&&&\\
\hline
& 3.46$\times 10$ & 2.46  & 2.53$\times 10^2$ & 2.90$\times 10^2$ & 1.37$\times 10$ & 7.38$\times 10^{-2}$\\
$y=0.0$ & 9.90  & -2.55$\times 10^{-1}$  & 1.77$\times 10^2$ & 1.87$\times 10^2$ & 1.22$\times 10$ & 1.03$\times 10^{-1}$ \\
& 3.10  & -2.10$\times 10^{-1}$  & 9.98$\times 10$ & 1.03$\times 10^2$ & 9.42  & 1.44$\times 10^{-1}$ \\
\hline
& 2.82$\times 10$ & 1.56  & 1.84$\times 10^2$ & 2.14$\times 10^2$ & 9.79  & 7.20$\times 10^{-2}$\\
$y=0.3$ & 7.48  & -2.15$\times 10^{-1}$  & 1.28$\times 10^2$ & 1.35$\times 10^2$ & 8.53  & 9.94$\times 10^{-2}$\\
& 2.25  & -1.54$\times 10^{-1}$  & 7.11$\times 10$ & 7.32$\times 10$ & 6.53  & 1.40$\times 10^{-1}$ \\
\hline
& 1.25$\times 10$ & 2.79$\times 10^{-1}$  & 5.97$\times 10$ & 7.24$\times 10$ & 3.06  & 6.63$\times 10^{-2}$\\
$y=0.6$ & 2.62  & -9.05$\times 10^{-2}$ & 4.01$\times 10$ & 4.26$\times 10$ & 2.42  & 8.94$\times 10^{-2}$\\
& 6.87$\times 10^{-1}$  & -4.69$\times 10^{-2}$ & 2.14$\times 10$ & 2.20$\times 10$ & 1.80  & 1.29$\times 10^{-1}$ \\
\hline
& 1.14  & -7.02$\times 10^{-3}$ & 3.71  & 4.85  & 1.69$\times 10^{-1}$  & 5.48$\times 10^{-2}$\\
$y=0.9$ & 1.47$\times 10^{-1}$  & -5.24$\times 10^{-3}$ & 2.15  & 2.29  & 1.04$\times 10^{-1}$  & 7.11$\times 10^{-2}$\\
& 2.87$\times 10^{-2}$ & -1.74$\times 10^{-3}$ & 1.02  & 1.05  & 7.18$\times 10^{-2}$ & 1.08$\times 10^{-1}$ \\
\hline
\end{tabular}

\begin{tabular}[c]{|c|p{2.4cm}|p{2.4cm}|p{2.4cm}|p{2.4cm}|p{2.4cm}|p{2.0cm}|}
\hline
GSI&&&&&&\\
$S=216.4\gev^2$&  $\Sigma_{gg}(pb/\gev^2)$     &  $\Sigma_{qg}(pb/\gev^2)$     &  $\Sigma_{q\bar{q}}(pb/\gev^2)$   &  $\Sigma(pb/\gev^2)$  & $\Sigma_T(pb/\gev^2)$&
$A_{TT}$\\
$E_{1\perp}=5\gev$&&&&&&\\
\hline
& 3.33$\times 10^{-1}$  & 4.43$\times 10^{-3}$ & 5.21  & 5.55  & 5.03$\times 10^{-1}$  & 1.43$\times 10^{-1}$ \\
$y=0.0$ & 7.56$\times 10^{-2}$ & -2.86$\times 10^{-3}$ & 3.47  & 3.55  & 3.73$\times 10^{-1}$  & 1.65$\times 10^{-1}$ \\
& 1.97$\times 10^{-2}$ & -1.38$\times 10^{-3}$ & 1.87  & 1.89  & 2.06$\times 10^{-1}$  & 1.72$\times 10^{-1}$ \\
\hline
& 2.06$\times 10^{-1}$  & 4.62$\times 10^{-4}$ & 2.86  & 3.07  & 2.61$\times 10^{-1}$  & 1.34$\times 10^{-1}$ \\
$y=0.3$ & 4.13$\times 10^{-2}$ & -1.64$\times 10^{-3}$ & 1.84  & 1.88  & 1.85$\times 10^{-1}$  & 1.55$\times 10^{-1}$ \\
& 1.00$\times 10^{-2}$ & -7.00$\times 10^{-4}$ & 9.62$\times 10^{-1}$  & 9.71$\times 10^{-1}$  & 9.89$\times 10^{-2}$ & 1.60$\times 10^{-1}$ \\
\hline
& 2.33$\times 10^{-2}$ & -3.38$\times 10^{-4}$ & 2.53$\times 10^{-1}$  & 2.76$\times 10^{-1}$  & 1.88$\times 10^{-2}$ & 1.07$\times 10^{-1}$ \\
$y=0.6$ & 3.08$\times 10^{-3}$ & -1.19$\times 10^{-4}$ & 1.37$\times 10^{-1}$  & 1.40$\times 10^{-1}$  & 1.09$\times 10^{-2}$ & 1.22$\times 10^{-1}$ \\
& 5.94$\times 10^{-4}$ & -3.73$\times 10^{-5}$ & 6.45$\times 10^{-2}$ & 6.50$\times 10^{-2}$ & 5.20$\times 10^{-3}$ & 1.26$\times 10^{-1}$ \\
\hline
\end{tabular}

\begin{tabular}[c]{|c|p{2.4cm}|p{2.4cm}|p{2.4cm}|p{2.4cm}|p{2.4cm}|p{2.0cm}|}
\hline
GSI&&&&&&\\
$S=216.4\gev^2$&  $\Sigma_{gg}(pb/\gev^2)$     &  $\Sigma_{qg}(pb/\gev^2)$     &  $\Sigma_{q\bar{q}}(pb/\gev^2)$   &  $\Sigma(pb/\gev^2)$  & $\Sigma_T(pb/\gev^2)$&
$A_{TT}$\\
$E_{1\perp}=6\gev$&&&&&&\\
\hline
& 8.39$\times 10^{-4}$ & -1.94$\times 10^{-5}$ & 3.20$\times 10^{-2}$ & 3.28$\times 10^{-2}$ & 3.90$\times 10^{-3}$ & 1.87$\times 10^{-1}$ \\
$y=0.0$ & 1.24$\times 10^{-4}$ & -5.61$\times 10^{-6}$ & 1.66$\times 10^{-2}$ & 1.68$\times 10^{-2}$ & 2.15$\times 10^{-3}$ & 2.02$\times 10^{-1}$ \\
& 2.42$\times 10^{-5}$ & -1.69$\times 10^{-6}$ & 7.73$\times 10^{-3}$ & 7.76$\times 10^{-3}$ & 1.01$\times 10^{-3}$ & 2.04$\times 10^{-1}$ \\
\hline
& 2.06$\times 10^{-4}$ & -4.77$\times 10^{-6}$ & 7.32$\times 10^{-3}$ & 7.52$\times 10^{-3}$ & 7.96$\times 10^{-4}$ & 1.66$\times 10^{-1}$ \\
$y=0.3$ & 2.54$\times 10^{-5}$ & -1.06$\times 10^{-6}$ & 3.44$\times 10^{-3}$ & 3.46$\times 10^{-3}$ & 3.92$\times 10^{-4}$ & 1.78$\times 10^{-1}$ \\
& 4.92$\times 10^{-6}$ & -2.95$\times 10^{-7}$ & 1.50$\times 10^{-3}$ & 1.51$\times 10^{-3}$ & 1.72$\times 10^{-4}$ & 1.79$\times 10^{-1}$ \\
\hline
\end{tabular}

\section{Notes for mathematica files}\label{sec:notes}
\begin{itemize}
\item{}\verb+hdTree+: $h_d^{(0)}(\tau_1,\rho)$;
\item{}\verb+hdLoop+: $h_d^{(1)}(\tau_1,\rho)$;
\item{}\verb+hpLarge+: $h_p^{(1)}(\tau_x,\tau_1,\rho)$;
\item{}\verb+hpSmall+: $h_p^{(1)}$ with $\tau_x$ expanded to $O(\tau_x^4)$;
\item{}\verb+hL+: $h_l^{(1)}(\tau_x,\tau_1,\rho)$.
\end{itemize}
Some parameters are introduced to give results in different schemes. In $\msb$-scheme,
\begin{align}
\text{tep}=1,\ \text{tct}=0,\ \text{nc}=\text{nb}=1,\ \text{nF}=3.
\end{align}
In zero-momentum subtraction of \cite{Nason:1989zy}, for bottom production,
\begin{align}
\text{tep}=1,\ \text{tct}=1,\ \text{nc}=0,\ \text{nb}=1,\ \text{nF}=4.
\end{align}
and for charm production,
\begin{align}
\text{tep}=1,\ \text{tct}=1,\ \text{nc}=1,\ \text{nb}=0,\ \text{nF}=3.
\end{align}
Other parameters are common, which are color factors and kinematical variables:
\begin{align}
N_1=N_cC_F^2,\ N_2=C_A N_cC_F^2,\ N_3=N_c C_F,\ ncolor=N_c,\
\rho_b=\frac{4m_b^2}{s},\ \rho_c=\frac{4m_c^2}{s},\ L_\mu=\ln\frac{\mu^2}{m^2}.
\end{align}
An example is given for unpolarized hard coefficients.
\newpage

\bibliography{ref_ts2}

\end{document}